\documentclass[aps, pra, superscriptaddress, twocolumn, amsfonts, amsmath, amssymb, floatfix]{revtex4-2}
\usepackage{graphicx}
\usepackage{float}
\usepackage{sidecap}
\usepackage{dcolumn}
\usepackage{bm}
\usepackage{epsfig}
\usepackage{braket}
\usepackage{color}
\usepackage{amsmath}
\usepackage[colorlinks=true, letterpaper=true, pdfstartview=FitV, linkcolor=blue, citecolor=blue, urlcolor=blue]{hyperref}

\begin{document}
\preprint{APS/123-QED}

\title{Band-edge superfluid of Bose-Einstein condensates in the spin-orbit-coupled Zeeman lattice}
 
\author{Huaxin He}
\affiliation{Institute for Quantum Science and Technology, Department of Physics, Shanghai University, Shanghai 200444, China}

\author{Fengtao Pang}
\affiliation{Institute for Quantum Science and Technology, Department of Physics, Shanghai University, Shanghai 200444, China}

\author{Hao Lyu}
\affiliation{Quantum Systems Unit, Okinawa Institute of Science and Technology Graduate University, Onna, Okinawa 904-0495, Japan}

\author{Yongping Zhang}
\email{yongping11@t.shu.edu.cn}
\affiliation{Institute for Quantum Science and Technology, Department of Physics, Shanghai University, Shanghai 200444, China}

\begin{abstract}

Since the first experimental realization of Bose-Einstein condensates in a spin-orbit-coupled Zeeman lattice, 
a wide range of applications have been found in these systems. Here, we systematically study the ground-state phase diagram of the systems. {The spin-orbit-coupled Zeeman lattice generates multiple energy minima in the lowest Bloch band. These minima may locate at Brillouin zone edges and center, or inside the Brillouin zone. Interacting atoms condensing into these minima constitute ground-state quantum phases. Among these phases, the band-edge condensation states have no analogue in either spin-orbit-coupled homogeneous systems or lattice systems without the spin-orbit coupling. They exist in a very broad parameter regime of the spin-orbit-coupled Zeeman lattice system. We address superfluidity of the band-edge states by analyzing their elementary excitations and superfluid fraction. }

\end{abstract}

\maketitle

\section{Introduction}
\label{Introduction}

The creation of synthetic spin-orbit coupling in atomic Bose-Einstein condensates (BECs)~\cite{lin2011spin,huang2016experimental,wu2016realization} has provided a clean and highly controllable platform for exploring novel quantum states and simulating many-body physics~\cite{goldman2014light,zhai2015degenerate,zhang2016properties}. The spin-orbit coupling enables engineering of dispersion relation, which leads to a rich ground-state phase diagram with various exotic phases including stripe phase and plane-wave phase~\cite{PhysRevLett.105.160403,PhysRevLett.107.150403,PhysRevLett.108.010402,PhysRevLett.108.225301,PhysRevA.93.033648}. Experimental observations of these phases~\cite{hamner2014dicke,ji2014experimental,PhysRevLett.114.125301,li2017stripe} stimulate further studies on elementary excitations~\cite{PhysRevA.86.063621,PhysRevLett.110.235302,zheng2013properties,PhysRevA.90.063624,PhysRevLett.114.105301,PhysRevA.102.023327,chen2022elementary}, superfluidity~\cite{zhu2012exotic, PhysRevA.94.033635,PhysRevA.95.033618,PhysRevA.106.023302, PhysRevLett.118.145302}, and spin-involved dynamics~\cite{PhysRevLett.109.115301,PhysRevA.86.041604,PhysRevA.88.021604,PhysRevA.90.013616,PhysRevLett.115.180402,PhysRevA.91.043604,qu2017spin,li2019spin1,PhysRevA.90.013616,PhysRevLett.130.156001,Hasan2022,PhysRevA.108.053316}.  

Loading interacting Bose atoms with the synthetic spin-orbit coupling into optical lattices~\cite{PhysRevLett.114.070401,wu2016realization} becomes an important platform to investigate many-body correlations~\cite{PhysRevA.85.061605,PhysRevLett.109.085303,PhysRevLett.109.085302}. In the Mott-insulating regime, spin-orbit coupling can generate many exotic magnetic phases~\cite{PhysRevA.85.061605,PhysRevB.90.085117,gong2015dz}, and magnetic phase transitions are identified by spin-spin correlations~\cite{PhysRevB.100.224420}. In the superfluid regime, superfluid phases still feature magnetic orders~\cite{PhysRevLett.109.085302}. The transitions between Mott insulators and superfluid phases have been studied~\cite{PhysRevA.96.061603}. In the presence of periodic potentials, spin-orbit coupling may make the lowest Bloch band flat~\cite{PhysRevA.87.023611,PhysRevA.91.023629}, and there may exist multiple energy minima in the lowest Bloch band~\cite{PhysRevA.94.043602}. How Bose atoms condense in the flat band~\cite{PhysRevLett.112.110404,PhysRevA.95.033603} or into Bloch states of multiple energy minima~\cite{PhysRevA.93.013601, PhysRevA.94.043629,martone2017quantum1,PhysRevA.94.043619,PhysRevA.102.033328,PhysRevA.102.053318} becomes a fascinating topic. Based on the ground-state phases of BECs with spin-orbit-coupled optical lattices, elementary excitations are investigated~\cite{PhysRevA.104.023311,10.1088/1367-2630/ad98b5}, and Bloch oscillations~\cite{PhysRevLett.117.215301,PhysRevA.99.023604,zhang2022bloch} and superfluidity breakdown~\cite{PhysRevA.89.061605,PhysRevA.103.063324} show interesting features.   By incorporating disorder in optical lattices, spin-orbit coupling induces an anomalous velocity, which leads to a random spatial separation between spin components~\cite{PhysRevLett.115.180402,PhysRevA.91.043604}, and the effect of spin-orbit coupling on Anderson localization is identified~\cite{PhysRevLett.118.105301,HuanZhang70305,PhysRevA.107.033312}. In experiments, optical lattices can induce the coupling between two energy minima in the lowest band of spin-orbit-coupled dispersion, which gives rise to experimentally observable long-lived superfluid stripe states~\cite{PhysRevA.99.051602}, and can excite the momentum-space Josephson dynamics~\cite{PhysRevLett.132.233403}.

In 2012, a landmark experiment realized a specific spin-dependent lattice in BECs by combining radio frequency (RF) and optical-Raman coupling fields~\cite{PhysRevLett.108.225303}. Unlike conventional optical lattices, this one-dimensional spin-dependent lattice (called Zeeman lattice) is created without using optical standing waves and naturally possesses the spin-orbit coupling with the linear momentum along the same direction of the lattice. Therefore, it is referred to as the spin-orbit-coupled Zeeman lattice (SOC ZL). Immediately after the invention, it was implemented in Fermi gases~\cite{PhysRevLett.109.095302}. The SOC ZLs have crucial applications in geometric pumping~\cite{PhysRevLett.116.200402} and Floquet engineering~\cite{PhysRevLett.129.040402}. Recently, it was experimentally demonstrated that the SOC ZLs restore Galilean invariance which is typically broken in conventional SOC systems~\cite{ome2024galilean}. Consequently, the SOC ZLs are versatile for manipulating SOC atoms by equivalently tuning the lattices. The exotic lattices were theoretically proposed to be realized by using pulsed gradient magnetic fields~\cite{LuoXY_2015}. The advantage of gradient magnetic field realizations is that the implemented SOC ZLs can be easily generalized to two-dimensional~\cite{Suref_2015} and to have rich lattice geometries~\cite{PhysRevA.95.013409}.  A further theory proposed to realize SOC ZLs by employing two independent optical-Raman couplings~\cite{PhysRevA.100.053606}. The interesting feature of resultant SOC ZLs is that the direction of the linear momentum in spin-orbit coupling can be perpendicular to the lattice direction.  Even though BECs in the SOC ZLs have been a prototypical model to study nonlinear phenomena~\cite{PhysRevLett.117.215301,PhysRevA.98.061601,PhysRevA.105.063323,PhysRevE.109.064219} and have found a broad range of applications~\cite{PhysRevLett.116.200402,PhysRevLett.129.040402,ome2024galilean,PhysRevA.95.013409}, their ground-state phase diagram remains unexplored.

In this paper, we systematically study the ground-state phase diagram of BECs in the SOC ZLs.  First, we determine the single-particle phase diagram in the parameter space of the Raman coupling strength and the depth of the Zeeman lattice. The diagram is identified based on the energy minimum of the lowest Bloch band. Second, we analyze how interacting atoms condense into the single-particle phase diagram by using a variational wave function. Condensed quantum phases constitute the ground-state phase diagram. We find that band-edge states in the ground-state phase diagram are of particular interest. Their superfluidity is further identified by elementary excitations and superfluid fraction.   The two lowest excitation bands of the band-edge states are featured with the gapless Goldstone mode and the gaped pseudo-Goldstone mode, and there is an energy avoided crossing between them. The nonzero superfluid fraction of the band-edge states provides another unique character showing that the states are different from Bloch states at Brillouin zone edges in lattice systems without spin-orbit coupling. {The band-edge states are unique because of their existence only in spin-orbit-coupled lattice systems.  They do not have analogues either in lattice systems or in SOC homogeneous systems. In the SOC ZLs, the band-edge states can exist in a very broad parameter regime, which is in favor of experimental observations. }

The remainder of this paper is organized as follows. In Sec.~\ref{MODEL}, we provide the theoretical model of BECs in the SOC ZL. In Sec.~\ref{Ground}, the ground-state phase diagram is identified. For this purpose, we first analyze the single-particle phase diagram in parameter space in Sec.~\ref{single}, and then study the ground-state phase diagram using variational method in  Sec.~\ref{variation}. In Sec.~\ref{band-edge}, we address properties of the band-edge states by showing their elementary excitations in  Sec.~\ref{elementary excitations} and superfluid fraction in  Sec.~\ref{Superfluid}. Finally, the conclusion follows
in Sec.~\ref{conclusion}.

\begin{figure*}[t]
	\includegraphics[ width=0.98\textwidth]{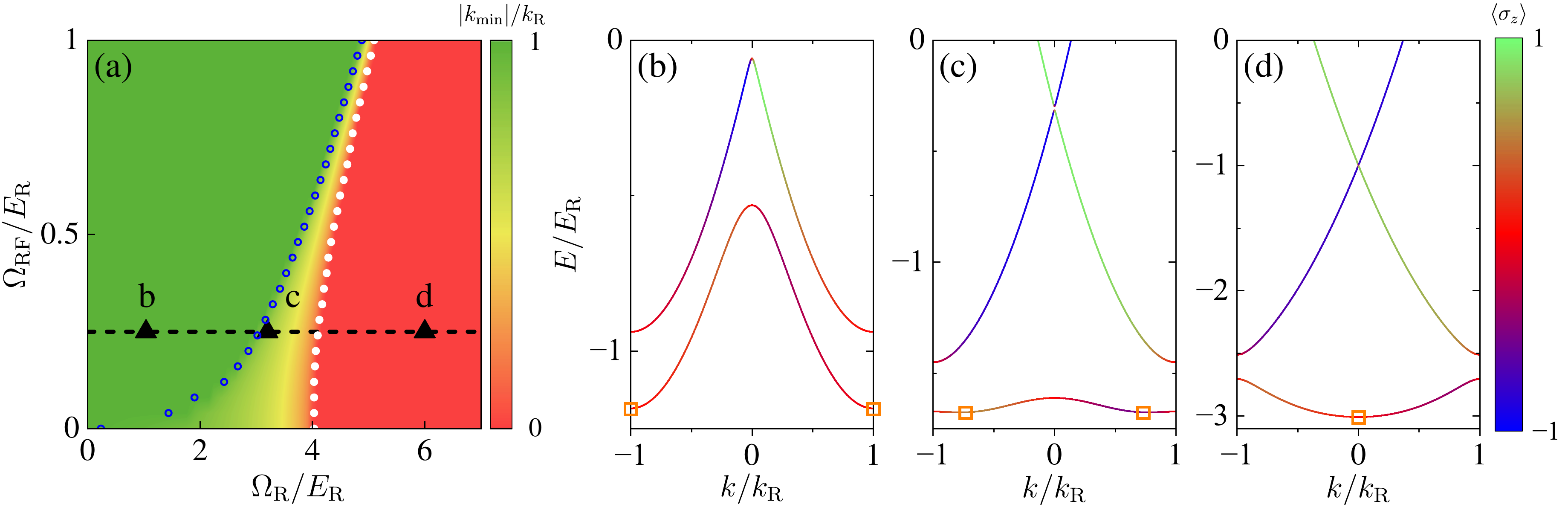}
	\caption{Phase diagram and Bloch spectrum of the single-particle 
        Hamiltonian $H_\text{sin}$ of the SOC ZL.
         (a) The single-particle phase diagram in the parameter space $(\Omega_{\text{R}}, \Omega_{\text{RF}})$. The diagram is identified according to values of $|k_\text{min}|$ which is the location of the energy minimum in the lowest Bloch band of $H_\text{sin}$. The green-colored region represents the band-edge minima $|k_\text{min}|/k_\text{R}=1$ with the edges at $k/k_\text{R}=\pm1$. Its phase boundary is shown by blue-open circles. The red-colored region corresponds to the band-center minimum, $k_\text{min}=0$, with the boundary shown by white-solid circles which may be approximated as $\Omega_\text{R}/E_\text{R}=4$. In the particular area surrounded by the open and solid circles, there are two energy minima in the lowest band located at $\pm |k_\text{min}|$ with $|k_\text{min}|/k_\text{R}<1$.  The black-dashed line marks $\Omega_{\text{RF}}/E_{\text{R}}=0.25$. (b)-(d) Bloch spectra of $H_\text{sin}$ with the parameters corresponding to the triangles labeled by ``b", ``c" and ``d", respectively. The spectrum is incorporated with the color scale representing value of the spin polarization $\langle\sigma_z\rangle$ of associated Bloch waves. The energy minima in the lowest Bloch band are represented by squares. 
         }
	\label{fig1}
\end{figure*}

\section{Model of BECs in the SOC ZL}
\label{MODEL}


We consider two hyperfine states of $^{87}$Rb atoms, which are simultaneously coupled by a pair of Raman lasers and a RF field~\cite{PhysRevLett.108.225303,ome2024galilean}. These couplings are described by the Hamiltonian:
\begin{equation}
	\label{Coupling}
        H= \frac{p_x^2}{2m} + \begin{pmatrix}
        -\frac{\Delta \epsilon}{2} & \mathcal{C} \\
        \mathcal{C}^* & \frac{\Delta \epsilon}{2}
        \end{pmatrix},
\end{equation}
with 
\begin{equation}
	\mathcal{C}=\frac{\Omega_\text{R}}{2} e^{ i2k_\text{R} x + i\Delta\omega_\text{R} t  }  +\frac{\Omega_\text{RF}}{2}e^{i\omega_{\text{RF}} t }. 
\end{equation}
Here, $p_x$ is the momentum of atoms along the propagation direction of the Raman laser, $m$ is the atomic mass, and $\Delta \epsilon$ is the energy difference between two hyperfine states. They are coupled by a two-photon Raman transition induced by the two Raman lasers with the coupling strength $   \Omega_\text{R} $.  $ k_\text{R} $ is the wave number of the Raman lasers, and $ \Delta\omega_\text{R}$ is their frequency difference.  The Raman lasers propagate oppositely so that there is a momentum transfer $2\hbar k_\text{R}$ between atoms and the Raman lasers during the coupling process~\cite{zhang2016properties}.  Simultaneously, the two hyperfine states are also coupled by the RF field with the frequency $\omega_{\text{RF}}$, leading to a coupling strength of $ \Omega_\text{RF}$.  In experiments, the perfect match $\Delta \omega_\text{R}=\omega_{\text{RF}} $ is implemented~\cite{PhysRevLett.108.225303,PhysRevLett.109.095302,PhysRevLett.116.200402,PhysRevLett.129.040402,ome2024galilean}.  
Combination of the Raman-laser-induced coupling and the RF coupling leads to the  SOC ZL~\cite{PhysRevLett.108.225303,PhysRevLett.109.095302,PhysRevLett.116.200402,PhysRevLett.129.040402,ome2024galilean,LuoXY_2015,PhysRevA.95.013409,PhysRevA.100.053606}.
Its single-particle Hamiltonian is denoted by $H_\text{sin}$,
which can be transformed to be time-independent by applying the unitary transformation $U=e^{i( k_\text{R}x+\omega_{\text{RF}} t/2 )\sigma_z }$~\cite{lin2011spin},
\begin{align}
\label{singleHamiltonian}
 H_\text{sin}& = U^\dagger H U \notag -i\hbar U^\dagger \frac{\partial U}{\partial t}\\
& = \frac{p_{x}^{2}}{2 m}+ \frac{\hbar k_{\text{R}}}{m} p_{x} \sigma_{z}+\frac{ \delta}{2} \sigma_{z}+\frac{\Omega_{\text{R}}}{2} \sigma_{x}  \notag\\
& \phantom{={}}+\frac{ \Omega_\text{RF}}{2}\left[\cos \left(2 k_{\text{R}} x\right) \sigma_{x}+\sin \left(2 k_{\text{R}} x\right) \sigma_{y}\right].
\end{align}
where $\{ \sigma_x,  \sigma_y,  \sigma_z\}$ are standard Pauli matrices, and the spin-orbit coupling arises with a strength of $\hbar k_\text{R}/m$. 
The RF coupling effectively behaves as a ZL with the lattice depth $\Omega_\text{RF} /2$. $\delta=\hbar \omega_{\text{RF}} - \Delta \epsilon$ is the detuning with $\hbar$ being the reduced Planck constant. In experiments, $\hbar \omega_{\text{RF}} =\Delta \epsilon$ can be readily achieved ~\cite{PhysRevLett.108.225303,PhysRevLett.109.095302,ome2024galilean}, leading to the disappearance of the detuning, i.e., $\delta=0$, which is the case that we consider in the remainder of the work.

The behavior of BECs in the SOC ZL is described by the Gross-Pitaevskii equation (GPE)~\cite{lin2011spin,PhysRevLett.108.225303},
\begin{equation}
	\label{nonGPE}
	i \hbar \frac{\partial \Psi}{\partial t}=\left[H_{\text{sin}}+ H_{\text{non}} \right]\Psi.
\end{equation}
Here, $ \Psi = ( \Psi_1, \Psi_2)^T$ is the wave function of two hyperfine states, with the normalization condition in a unit cell, $\int_{\text{cell}}dx(|\Psi_1|^2+|\Psi_2|^2)=1$.
$H_{\text{non}}$ describes interactions originating from atomic collisions,
\begin{equation}
	\label{nonline}
	H_{\text{non}} = \left(\begin{array}{cc}	g_{11}\left|\Psi_1\right|^2+g_{12}\left|\Psi_2\right|^2 & 0 \\
		0 & g_{22}\left|\Psi_2\right|^2+g_{12}\left|\Psi_1\right|^2
	\end{array}\right),
\end{equation}
where the coefficients $g_{ij} = 2\hbar a_{ij}N\sqrt{\omega_{y}\omega_{z}}$ ($i,j =1,2$) are the interaction coefficients. {Here $N$ denotes the atom number inside a unit cell as we use the normalization condition.} $a_{ij}$ represents the characteristic lengths of $s$-wave scattering. The trapping frequencies $\omega_y$ and $\omega_z$  in the transverse directions are sufficiently large to ensure that the system is effectively reduced to a quasi-one-dimensional form.  $^{87}$Rb BEC exhibits nearly identical $s$-wave scattering lengths, satisfying $a_{11} \approx a_{22} \approx a_{12}$~\cite{PhysRevLett.108.225303,ome2024galilean}. Therefore, for the convenience of numerical calculations, we use the equal coefficients, $g_{11}=g_{22}=g_{12}=g$. All physical parameters are expressed in appropriate units. The units of energy, length, momentum and time are defined as the recoil energy $E_{\text{R}} = \hbar^2k_{\text{R}}^2/2m$, $k_\text{R}^{-1}$, $k_\text{R}$ and $\hbar/E_\text{R}$, respectively.  In details, in the following numerical calculations, we consider transverse trapping frequencies \((\omega_y, \omega_z) = 2\pi \times (160, 190)~\text{Hz}\)~\cite{ome2024galilean} and assume $N=60$. Under these conditions, the interaction coefficient of $^{87}$Rb BEC is given by $g k_\text{R}/E_\text{R} = 0.0848$.

\section{Ground-state Phase diagram}
\label{Ground}

\subsection{Single-particle phase diagram according to energy minima in the lowest Bloch band}
\label{single}

We systematically investigate the ground-state phase diagram of BECs in the SOC ZL, which is governed by the GPE. The phase diagram is strongly influenced by the dispersion relation of the single-particle Hamiltonian $H_\text{sin}$~\cite{PhysRevA.93.013601, PhysRevA.94.043629}. Therefore, we first analyze its dispersion.

In the absence of the ZL, i.e., when $\Omega_{\text{RF}}=0$, the single-particle Hamiltonian in momentum space takes the form $ H_\text{sin}=\hbar^2k^2/2m +\hbar^2k_\text{R} k \sigma_z /m +\Omega _\text{R} \sigma_x/2 $. Here, $\hbar k $ represents the quasimomentum. The dispersion relation of $ H_\text{sin}$ consists of two branches,
\begin{equation}
E_\pm(k)=\frac{\hbar^2k^2}{2m} \pm \sqrt{\left(\frac{\hbar^2k_\text{R}k}{m}\right)^2+\left(\frac{\Omega_\text{R}}{2}\right)^2}.
\end{equation}
By solving $\partial E_{-}/\partial k=0$, we find the energy minima located at $k_\text{m}= \pm k_\text{R} \sqrt{ 1-(\Omega_\text{R}/4E_\text{R})^2 } $ when $\Omega_\text{R}/E_\text{R} < 4$ or at $k_\text{m}= 0 $ when $\Omega_\text{R}/E_\text{R} \geqslant 4$. Therefore, depending on $\Omega_{\text{R}}$, there might be two energy minima or one minimum in the lower band $E_{-}(k)$. Atoms condense into these minima forming different quantum phases~\cite{PhysRevLett.108.225301}.

The presence of the ZL in $H_\text{sin}$ requires that single-particle solutions be Bloch states~\cite{PhysRevA.65.025601}, $\Psi(x)=e^{ikx }\Phi(x)$ with  $ \Phi = ( \Phi_1, \Phi_2)^T$ being a periodic function with the same periodicity as the ZL. By expanding $\Phi(x)$ in a plane-wave basis,  $H_\text{sin}$ can be diagonalized. The resulting dispersion $E(k)$ features Bloch band-gap spectrum.  The ZL provides the atoms with crystal momentum $2\hbar k_\text{R}$ which translates $E_{-}(k)$ to  $E_{-}(k+2n k_\text{R})$ with $n=\pm 1, \pm 2,\cdots$.  There is a possible energy crossing between $E_{-}(k)$ and  $E_{-}(k+2 k_\text{R})$ at $k/ k_\text{R}=-1$, and between  $E_{-}(k)$ and  $E_{-}(k-2 k_\text{R})$ at $k/k_\text{R}=1$. The ZL further breaks these energy degeneracies by opening an energy gap and the gap size closely depends on the lattice depth $\Omega_\text{RF}$. The remaining part of  $E_{-}(k)$ together with the resulting structures induced by gap-opening at $k/k_\text{R}=\pm 1$ form the lowest Bloch band inside the first Brillouin zone $-1 \leqslant k/k_\text{R} \leqslant 1$~\cite{PhysRevA.87.023611}. 

We identify the location of the energy minima, $k_\text{min}$, in the lowest Bloch band. The phase diagram based on values of $|k_{\text{min}} |$ in the parameter space $(\Omega_{\text{R}}, \Omega_{\text{RF}})$ is shown in Fig.~\ref{fig1}(a), which is divided into three regions, each labeled with a different color. 

{\subsubsection{band-edge minima} }
 In the green-colored region, the energy minima of the lowest Bloch band are located at the edges of the Brillouin zone, $ |k_{\text{min}} | /k_\text{R} =1 $.  A typical spectrum including two lowest bands with the parameters labeled by the triangle ``b" is illustrated in Fig.~\ref{fig1}(b), which is incorporated with the color scale representing values of the spin polarization $\langle \sigma_z \rangle $ of associated Bloch states.  $\langle \sigma_z \rangle = N_1 - N_2$, where
$N_i=\int_\text{cell}dx |\Psi_i(x)|^2$ is the population of the $i$-th components inside a unit cell of the ZL. Bloch states in the lowest band are almost unpolarized, $\langle \sigma_z \rangle = 0$. There are two energy minima at the Brillouin zone edges $ k_{\text{min}}  /k_\text{R} =\pm 1 $ denoted by squares.  Considering the crystal-momentum $2\hbar k_\text{R}$, Bloch states of band-edge minima are equivalent. They are the same state with the finite quasimomentum at the band edges, which are named the band-edge state. Despite having finite quasimomentum, the band-edge states conserve time-reversal symmetry
\begin{equation}
\label{Time}
    \mathcal{T}=\mathcal{K} \sigma_x,
\end{equation}
which leads to $\langle \sigma_z \rangle = 0$. Here, $\mathcal{K}$ is the complex conjugate operator.

Figure~\ref{fig1}(a) shows the band-edge minima exist over a very wide region in the  $(\Omega_{\text{R}}, \Omega_{\text{RF}})$ space. When $\Omega_\text{R}$ is very small, the energy minima of $E_{-}(k)$ are located at $k_\text{m}/k_\text{R} \approx \pm 1 $. The ZL lowers the energy at the edges $ \pm k_\text{R}$ and opens gaps. Combining the energy minima of $E_{-}(k)$ and the band-edge energy pushed down naturally generates the global minima at the edges. By increasing $\Omega_\text{R}$,  $k_\text{m} $ will deviate from the edges towards $0$. A large depth $\Omega_\text{RF}$ is needed to push down the energy at the edges to be lower than the minima of $E_{-}(k)$. Therefore, there exists a critical $\Omega_\text{RF}$ beyond which the band-edge minima appear. Such critical values are demonstrated by blue-open circles in  Fig.~\ref{fig1}(a). The dependence of the critical values on $\Omega_\text{R}$ looks like a parabolic shape, which consequently causes the band-edge minima to exist in a wide area. This wide existence becomes the main feature of the SOC ZL and is different from the case of the usual SOC optical lattices where the band-edge minima exist in a limited parameter regime~\cite{PhysRevA.93.013601, PhysRevA.94.043629}. We emphasize that the physical reason for the wide existence lies in the perfect match between the crystal-momentum $2\hbar k_\text{R}$ provided by the ZL and the momentum scale $2\hbar | k_\text{m}|=2\hbar k_\text{R}$ due to the minima of $E_-$ in the absence of $\Omega_\text{R}$. 

{\subsubsection{band-center minimum} }
In the red-colored region in Fig.~\ref{fig1}(a), the lowest Bloch band has a single energy minimum located at the band center $ k_{\text{min}}=0$. A typical spectrum with the parameters labeled by  the triangle ``d"  in Fig.~\ref{fig1}(a) is demonstrated in Fig.~\ref{fig1}(d).  The lowest Bloch band resembles the usual optical-lattice-induced dispersion. The energy minimum represented by the square features unpolarization $\langle \sigma_z \rangle = 0$. The Bloch state of the band-center minimum is referred to as the band-center state which conserves time-reversal symmetry $\mathcal{T}$. In the absence of 
the ZL ($\Omega_\text{RF}=0$), the energy minimum occurs at $k=0$ when $\Omega_\text{R}/E_\text{R}> 4$ in $E_-$. The presence of the ZL does not significantly modify the critical value $\Omega_\text{R}/E_\text{R}=4$ for the existence of the band-center minimum, as shown by the white-solid circles in Fig.~\ref{fig1}(a). Physically, this is reasonable, as the ZL primarily affects the band edges.

{\subsubsection{double minima} }

A distinct region exists between the green and red areas in Fig.\ref{fig1}(a), enclosed by open and solid circles. In this region, the lowest Bloch band has two energy minima located at $\pm |k_\text{min}|$ with $|k_\text{min}|/ k_\text{R}<1$. A typical spectrum with the parameters labeled by  the triangle ``c"  in Fig.~\ref{fig1}(a) is demonstrated in Fig.~\ref{fig1}(c). The two minima in the lowest band (denoted by the squares) possess opposite spin polarizations $\langle \sigma_z \rangle$. The Bloch state of each minimum breaks the time-reversal symmetry due to the finite quasimomentum leading to nonzero $\langle \sigma_z \rangle$.  The Bloch state at $-k_\text{min}$, $\Psi_{ -k_\text{min}}$, is time-reversal symmetrical with the Bloch state at $k_\text{min}$, i.e, $\Psi_{ -k_\text{min}}=  \mathcal{T} \Psi_{ k_\text{min}} $.  $|k_\text{min}|$ moves towards the edge ($k_\text{R}$) as decreasing $\Omega_\text{R}$ and moves towards the center ($k=0$) as increasing $\Omega_\text{R}$. The combination of the parabolic-like boundary [shown by the open circles in Fig.~\ref{fig1}(a)] and the approximating constant boundary $\Omega_{R}/E_\text{R}=4$ (shown by the white-solid circles) makes the area to be very narrow for large values of $\Omega_\text{RF}$.

\subsection{Ground-state phase diagram by a variational wave function}
\label{variation}

Ground-state quantum phases correspond to condensations into Bloch states of at the energy minima in the lowest Bloch band. Based on the previous discussion, the possible condensate states include the band-edge state, the band-center state, and the Bloch states associated with the two energy minima. All these condensate states can be approximately described by a variational wave function~\cite{PhysRevA.93.013601}, 
\begin{equation}
\label{ZL}
\Psi(x)=\begin{bmatrix}
\Psi_{1}(x) \\
\Psi_{2}(x)
\end{bmatrix}=e^{ik_\text{c}x } \psi_{+k_\text{c}}(x)+e^{-ik_\text{c}x} \psi_{-k_\text{c}}(x).
\end{equation}
It is a superposition of two Bloch states with quasimomentum $\pm k_\text{c}$.  The periodic function $\psi_{k}(x)$ is expanded in a plane-wave basis,
\begin{equation}
	\psi_{k}(x) =\sum_{n=-L}^{L}\left(\begin{array}{l}
		a_{n, k} \\
		b_{n, k}
		\end{array}\right) e^{i2nk_\text{R}x},
\end{equation}
with $L$ being the cutoff for plane-wave modes and $a_{n,k}$ and $b_{n,k}$ are expansion coefficients. The unknown parameters $k_\text{c}$, $a_{n,\pm k_\text{c}}$, and $b_{n,\pm k_\text{c}}$ in the variational wave function should be determined by minimizing the energy functional $\mathcal{E}$, which is associated with the GPE in Eq.~(\ref{nonGPE}),  
\begin{equation}
	\label{EF1}
	\begin{aligned}
		\mathcal{E}= \int d x\left\{\Psi^{\dagger} H_{\text{sin}} \Psi+\frac{g}{2}\sum_{i,j=1,2}\left|\Psi_{i}\right|^{2}\left|\Psi_{j}\right|^{2}\right\}.
	\end{aligned}
\end{equation}
By substituting Eq.~(\ref{ZL}) into Eq.~(\ref{EF1}), the minimization condition is given by $\partial \mathcal{E}/\partial X=0$, where $X=\{ k_\text{c}, a_{n,\pm k_\text{c}}, b_{n,\pm k_\text{c}} \}
$. Note that in the variational wave function, the superposition of the two Bloch states has equal probability amplitude. The occupation of each Bloch state has already been absorbed into the parameters $a_{n,\pm k_\text{c}}, b_{n,\pm k_\text{c}}$.

\begin{figure*}[t]
	\includegraphics[ width=1  \textwidth]{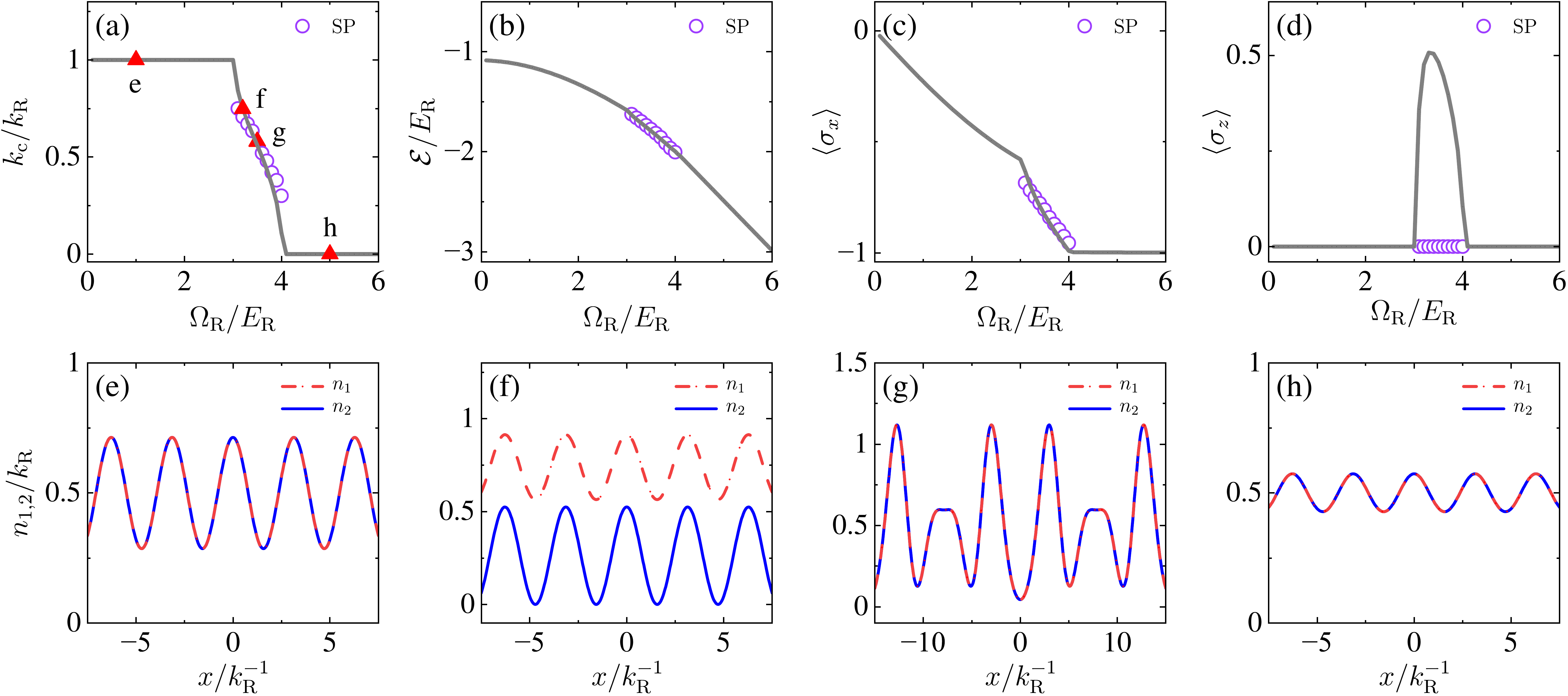}
	\caption{The ground-state phase diagram for a fixed depth of the ZL 
         $\Omega_{\text{RF}}/E_{\text{R}} = 0.25$. The interaction coefficient is $g k_\text{R}/E_\text{R} = 0.02$. (a) The condensed quasimomentum $k_\text{c}$ as a function of the Raman coupling $\Omega_{\text{R}}$. $k_\text{c}/k_\text{R}=1$ is in the band-edge phase, $k_\text{c}/k_\text{R}=0$ is in the band-center phase, and in-between is the single Bloch phase. In (b)-(d), the energy $\mathcal{E}$, the polarizations, $\langle \sigma_x \rangle$ and $\langle \sigma_z \rangle$ are shown as a function of $\Omega_{\text{R}}$ respectively.  In (a)-(d), the stripe solutions are shown by circles. (e) Density distributions $n_\text{1,2}(x)=
         |\Psi_{1,2}|^2$ of a band-edge state with $\Omega_{\text{R}}/E_{\text{R}} = 1$. (f) Density distributions of a single Bloch state with $\Omega_{\text{R}}/E_{\text{R}} = 3.2$. (g) Density distributions of a stripe state with $\Omega_{\text{R}}/E_{\text{R}} = 3.5$. (h) Density distributions of a band-center state with $\Omega_{\text{R}}/E_{\text{R}} = 5$. The parameters of (e)-(h) corresponds to the triangles labeled by ``e"-``h" in (a), respectively.}
	\label{fig2}
\end{figure*}

Quantum phases are classified based on the resultant parameters obtained from the minimization. Furthermore, we characterize the quantum phases in terms of two distinct spin polarizations,
\begin{equation}
	\label{Eq8}
	\left\langle\sigma_{x , z}\right\rangle= \int_\text{cell} \Psi^{\dagger} \sigma_{x , z} \Psi d x.
\end{equation}
If the quantum phase preserves the time-reversal symmetry $\mathcal{T}$ in Eq.~(\ref{Time}), then it satisfies $\langle \sigma_z \rangle = 0$.  According to the Hellmann-Feynman theorem, we obtain $\partial\mathcal{E}/\partial \Omega_\text{R}=  \langle\partial H_\text{sin}/ \partial \Omega_\text{R} \rangle =\langle \sigma_x \rangle/2 $. Consequently, by analyzing the continuity of $\langle \sigma_x \rangle$ as a function of $\Omega_\text{R}$, we can determine the order of the phase transition.

(i) If the calculated result yields $k_\text{c}=k_\text{R}$, the corresponding quantum phase is identified as the band-edge state. Its spin polarization $\langle \sigma_z \rangle=0$ confirms that it preserves the time-reversal symmetry $\mathcal{T}$. The band-edge phase emerges within the parameter region where the band-edge minima appear, as shown in Fig.~\ref{fig1}(a) (the green-colored region).   

(ii)  If the calculated result yields $k_\text{c}=0$, the corresponding phase is identified as the band-center state. This phase preserves time-reversal symmetry, as indicated by its spin polarization $\langle \sigma_z \rangle=0$. The band-center phase emerges within the parameter region where the band-center minimum appears, as shown in Fig.~\ref{fig1}(a) (the red-colored region). In this regime, the Raman coupling $\Omega_\text{R}$ dominates, resulting in $|\langle \sigma_x \rangle|=1$.

(iii) When the calculated results indicate that $k_\text{c}$ is finite but not equal to $k_\text{R}$, two distinct solutions emerge, which can be distinguished based on the values of $a_{n,\pm k_\text{c}}$ and $b_{n,\pm k_\text{c}}$.  One solution is characterized by the absence of either $\psi_{+k_\text{c}}(x)$ or $\psi_{-k_\text{c}}(x)$ in Eq.~(\ref{ZL}), and is referred to as the single-momentum Bloch state. Notably, both the band-edge and band-center states are special cases of the single-momentum Bloch state, with quasimomentum corresponding to the band edge or band center, respectively. Since $k_\text{c} \neq 0$, the single-momentum Bloch state breaks time-reversal symmetry $\mathcal{T}$, leading to a nonzero spin polarization, i.e., $\langle \sigma_z \rangle \neq 0$. 

(iv) The other solution features that both of $\psi_{+k_\text{c}}(x)$ and $\psi_{-k_\text{c}}(x)$ are nonzero and they are related by $\psi_{+k_\text{c}}(x)= \mathcal{T}\psi_{-k_\text{c}}(x)$. This solution is the equal superposition of two Bloch states that are correlated by $\mathcal{T}$. We call this solution the stripe state considering its density periodic modulation due to the mixing of two Bloch states. Consequently, the stripe state respects $\mathcal{T}$, leading to $\langle \sigma_z \rangle =0$. Both the single-momentum Bloch state and the stripe state may exist in the parameter region surrounded by open and solid circles in Fig.~\ref{fig1}(a).

Applying the above variational method and the classification of quantum phases, we study ground-state phase diagram and phase transitions by varying the Raman coupling $\Omega_\text{R}$ for a fixed lattice depth $\Omega_{\text{RF}}/E_\text{R}=0.25$ which corresponds to the horizontal dashed line in the parameter space $(\Omega_\text{R}, \Omega_{\text{RF}})$ in Fig.~\ref{fig1}(a).  The calculated $k_\text{c}$ is demonstrated as a function of $\Omega_\text{R}$ in Fig.~\ref{fig2}(a). In the region of $0<\Omega_\text{R}/E_\text{R}<3$, $ k_\text{c}/k_\text{R}=1$, so that the ground state is in the band-edge phase. Density distributions $n_{1,2}(x)=|\Psi_{1,2}(x)|^2$ of a typical band-edge state are shown in  Fig.~\ref{fig2}(e). Two components have the same period as the ZL, and $n_1(x)=n_2(x)$ indicates the time-reversal symmetry, so that  $\langle \sigma_z \rangle=0$, as shown in Fig.~\ref{fig2}(d).  In the region of  $3<\Omega_\text{R}/E_\text{R}<4$, $k_\text{c}$ continuously decreases from 1 to 0 as the increase of $\Omega_\text{R}$ [see the solid line in Fig.~\ref{fig2}(a)]. In this region, there are two energy minima in the lowest band of $H_\text{sin}$, the solution shown in  Fig.~\ref{fig2}(a) spontaneously chooses the one of them to condense. Therefore, the solution is in the single-momentum Bloch state phase.  Density distributions of a typical single-momentum Bloch state are demonstrated in Fig.~\ref{fig2}(f). The outstanding feature is $n_1(x) \neq n_2(x)$, indicating that there is no time-reversal symmetry and $\langle \sigma_z \rangle \neq 0$, which is confirmed in the results shown in Fig.~\ref{fig2}(d).  In the region of $3<\Omega_\text{R}/E_\text{R}<4$, we also find the stripe solutions, $ e^{ik_\text{c}x } \psi_{+k_\text{c}}(x)+ \mathcal{T}e^{ik_\text{c}x } \psi_{+k_\text{c}}(x)$.  The $k_\text{c}$ of the stripe states are shown by circles in Fig.~\ref{fig2}(a). They almost coincide with these of the single-momentum Bloch states. The energy of the stripe states is demonstrated in Fig.~\ref{fig2}(b). It is clear that the energy also almost coincides with that of the single-momentum Bloch states (shown by the solid line).  The coincidences are caused by the equal interacting coefficients we considered. Typical density distributions of a stripe state are shown in Fig.~\ref{fig2}(g), featuring  $n_1(x) = n_2(x)$, which is a signature of respecting $\mathcal{T}$.  The results of $\langle \sigma_z \rangle=0$ (shown by open circles) in  Fig.~\ref{fig2}(d) also confirm the symmetry $\mathcal{T}$. The most important feature of the stripe states is that their densities have multiple periods. When $\Omega_\text{R}/E_\text{R}>4$, we find the band-center ground state and $k_\text{c}=0$ as shown in Fig.~\ref{fig2}(a). Density distributions of a band-center state are described in Fig.~\ref{fig2}(h), from which it can be seen that $\mathcal{T}$ is conserved. In the band-center phase, $|\langle \sigma_x \rangle|$ is maximized as shown in Fig.~\ref{fig2}(c).  

In Fig.~\ref{fig2}(a), as increasing $\Omega_\text{R}$, the band-edge phase, the single-momentum Bloch state phase and the band-center phase successively appear. The boundaries between them are $\Omega_{\text{R}}/E_\text{R}\approx 3$ and $\Omega_{\text{R}}/E_\text{R}\approx 4$. Such boundaries match these in the single-particle phase diagram shown in Fig.~\ref{fig1}(a).  The reason of the match is that we consider {weak interactions and meanwhile  they have equal interacting coefficients}. It is known that interactions with non-equal coefficients can modify phase-transition boundaries in SOC homogeneous systems~\cite{PhysRevLett.108.225301}. Also due to the equal interactions, the stripe phase is almost degenerate with the single-momentum Bloch phase.
The energy $\mathcal{E}$ changes smoothly across different phases as shown in Fig.~\ref{fig1}(b). Furthermore, the resultant $\langle \sigma_x \rangle$ also evolves continuously as a function of $ \Omega_\text{R}$ in Fig.~\ref{fig1}(c). Meanwhile, there are sharp changes at the boundaries indicating the transitions are second order.
The second-order transitions are consistent with the transitions in SOC optical lattices~\cite{PhysRevA.94.043629}. 

{As shown in Fig.~\ref{fig2}(a), the band-edge state, the single-momentum state and the band-center state have different occupations in momentum space. This provides an experimental means to distinguish them by time-of-flight observations.  }

\begin{figure*}[t]
	\centerline{\includegraphics[ width=0.95\textwidth]{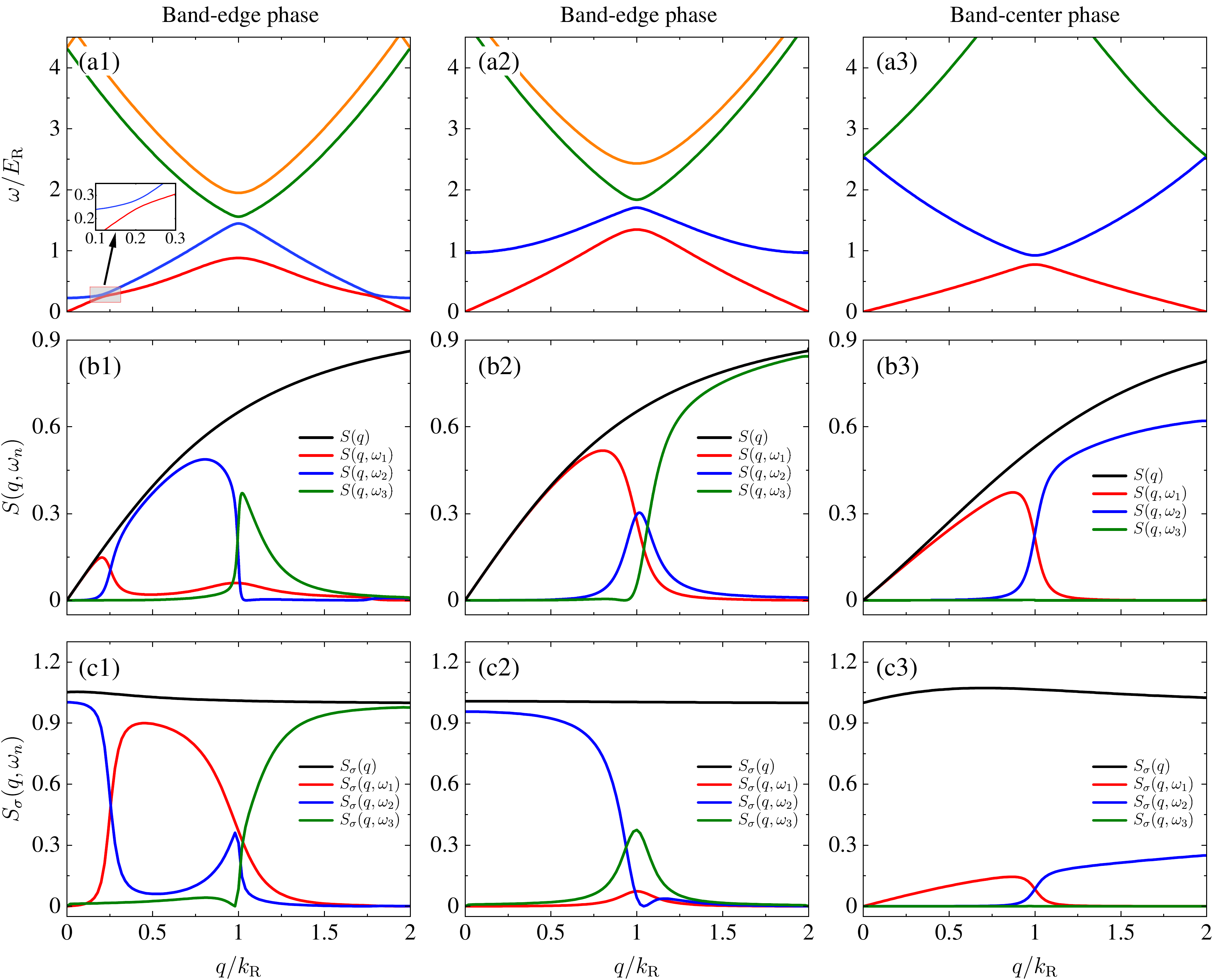}}
	\caption{Elementary excitation spectrum and static structure factor in the band-edge phase. The interaction coefficient is $g k_\text{R}/E_\text{R} = 0.0848$. (a1) Excitation spectrum of the band-edge state with $(\Omega_{\text{R}}, \Omega_{\text{RF}})/E_{\text{R}} = (1, 0.25)$.  Only the lower four bands are shown.  The inset is the zoom-in of the energy avoided crossing between the first and second bands.  (b1)  The static structure factor $S(q,\omega_n)$ calculated form the excitation spectrum shown in (a1). The dark line is total structure factor $S(q)=\sum_n S(q,\omega_n)$ by considering all excitation bands. (c1) The static structure factor $S_\sigma(q,\omega_n)$ calculated form the excitation spectrum shown in (a1). The dark line is total structure factor $S_\sigma(q)=\sum_n S_\sigma(q,\omega_n)$ by considering all excitation bands.    (a2), (b2) and (c2) show the same quantities as (a1), (b1) and (c1) for the band-edge states with parameters  $(\Omega_{\text{R}}, \Omega_{\text{RF}})/E_{\text{R}} = (1, 1)$.  (a3), (b3) and (c3) show the same quantities as (a1), (b1) and (c1) but for the band-center states with parameters  $(\Omega_{\text{R}}, \Omega_{\text{RF}})/E_{\text{R}} = (6, 0.25)$.  
         }
	\label{fig3}
\end{figure*}

\section{Band-edge superfluid}
\label{band-edge}

The single-momentum Bloch phase, the band-center phase and the stripe phase of the SOC ZL have the analogues in the SOC homogeneous systems. However, there is no corresponding for the band-edge states. Considering the existence of the band-edge states in a very broad region of parameters in the SOC ZL, we further examine their properties.

\subsection{Elementary Excitations of the band-edge states}
\label{elementary excitations}

In a SOC BEC, elementary excitations are of particular interest~\cite{chen2022elementary}. We study elementary excitations of the band-edge states. For this purpose, we first construct a general wave function as
\begin{equation}
	\label{Eq9}
\Psi(x,t)=e^{ikx-i\mu t}\left \{\begin{bmatrix}
    \phi_1(x)\\
    \phi_2(x)
\end{bmatrix}+\delta\psi(x,t)\right \}. 
\end{equation}
It includes a ground state $e^{ikx-i\mu t} (\phi_1(x),\phi_2(x))^T$ satisfying the GPE (with $\hbar k$ and $\mu$ being quasimomentum and chemical potential respectively) and a perturbation $\delta\psi(x,t) $. The band-edge state has $k/k_\text{R}=1$. The perturbation should be
\begin{equation}
\label{Eq10}
    \delta\psi(x,t)=\begin{bmatrix}
        u_1(x)\\
        u_2(x)
    \end{bmatrix}e^{-i\omega t}  +\begin{bmatrix}
       v_1^*(x)\\
       v_2^*(x)
    \end{bmatrix}e^{i\omega^*t}.
\end{equation}
Here, $\hbar\omega$ is the perturbation energy, and $u_{1,2}(x)$ and $v_{1,2}(x)$ are perturbation amplitudes, satisfying the normalization condition $\sum_{j=1,2}\int dx\left[|u_j(x)|^2-|v_j(x)|^2\right]=1$. Substituting Eq.~(\ref{Eq9}) into Eq.~(\ref{nonGPE}), and  retaining only linear contributions in $u_{1,2}$ and $v_{1,2}$, we obtain the Bogoliubov-de Gennes (BdG) equations
\begin{equation}
\label{Eq11}
\mathcal{M}\Phi=\omega\Phi,
\end{equation}
with
\begin{equation}
\label{Eq12}
  \mathcal{M}=\begin{pmatrix}
        \mathcal{A}&\mathcal{B}\\
        \mathcal{-B^*}&\mathcal{-A^*}    \end{pmatrix}, \qquad \Phi=(u_1,u_2,v_1,v_2)^T.
\end{equation}
The matrices $\mathcal{A}$ and $\mathcal{B}$ are
\begin{equation}
\label{Eq13}
    \begin{aligned}
        \mathcal{A}&=\frac{\Omega_{\text{R}}}{2}\sigma_x+\frac{\Omega_{\text{RF}}}{2}\left[\cos(2k_\text{R}x)\sigma_x+\sin(2k_\text{R}x)\sigma_y\right]\\
        &\phantom{={}}+\begin{pmatrix}
            \mathcal{H}_1-\frac{i\hbar^2k_\text{R}}{m}(\partial_x+ik)&g\phi_1\phi_2^*\\
        g\phi_1^*\phi_2&\mathcal{H}_2+\frac{i\hbar^2k_\text{R}}{m}(\partial_x+ik)
        \end{pmatrix},\\
        \mathcal{B}&=g\begin{pmatrix}
            \phi_1^2&\phi_1\phi_2\\
            \phi_1\phi_2&\phi_2^2
        \end{pmatrix},
    \end{aligned}
\end{equation}
with
\begin{equation}
\label{Eq14}
    \begin{aligned}
        \mathcal{H}_1=-\frac{\hbar^2}{2m}\left(\frac{\partial}{\partial x}+ik\right)^2+2g  |\phi_1|^2+g|\phi_2|^2-\mu,\\
     \mathcal{H}_2=-\frac{\hbar^2}{2m}\left(\frac{\partial}{\partial x}+ik\right)^2+g|\phi_1|^2+2g|\phi_2|^2-\mu.
    \end{aligned}
\end{equation}
Since the band-edge states are spatially periodic, the BdG matrix $\mathcal{M}$ is periodic with the same period as the ZL. Therefore, the solutions of the BdG equations in Eq.~(\ref{Eq11}) should be Bloch states, $\Phi(x)=\Phi_q(x)e^{iqx}$, with $\hbar q$ being the quasimomentum of perturbation. The periodic function is 
\begin{equation}
\label{pwave}
    \Phi_q(x)=(u_{1q}(x),u_{2q}(x), v_{1q}(x),v_{2q}(x))^T.
\end{equation}
By expanding the periodic function in a plane-wave basis, the BdG equations can be diagonalized and elementary excitations $\omega (q)$ can be derived. 

A similar band-edge state in a SOC optical lattice system has been experimentally observed in Ref.~\cite{PhysRevA.99.051602}. Its elementary excitations have been theoretically revealed to relate with the pseudo-Goldstone mode in Ref.~\cite{PhysRevA.104.023311}. In Fig.~\ref{fig3}(a1), we show elementary excitations of the band-edge state with parameters  $(\Omega_{\text{R}}, \Omega_{\text{RF}})/E_{\text{R}} = (1, 0.25)$ in the SOC ZL. The excitation spectrum $\omega(q)$ features Bloch band-gap structures with the first Brillouin zone $-1\leqslant q/k_\text{R} \leqslant 1$.  In the long-wavelength limit ($q\rightarrow 0$), the first band is gapless, which originates from the gauge symmetry breaking. The second band is gaped, which is called the pseudo-Goldstone mode in Ref.~\cite{PhysRevA.104.023311}. The most outstanding characteristic of these two bands is that there is an energy avoided crossing between them at a finite momentum [see the zoom-in in Fig.~\ref{fig3}(a1)]. In the SOC optical lattice system, similar avoided crossing has an effect on the dynamics in momentum space~\cite{PhysRevLett.120.120401,PhysRevLett.132.233403}. 

Elementary excitations can be experimentally detected by Bragg spectroscopy.  Excitation efficiency of the spectroscopy, which is experimentally observable, is proportional to the static structure factor~\cite{PhysRevLett.114.105301}, which is defined as~\cite{chen2022elementary,pitaevskii2016bose}:
\begin{equation}
\label{Eq15}
\begin{aligned}
S(q,\omega_n)&=\left|\int\mathrm{d}x \left[ 
  \mathcal{P}  (\mathbb{I} \otimes \mathbb{I})  \Phi^n_q\right]\right|^2,\\
S_\sigma(q,\omega_n)&=\left|\int\mathrm{d}x \left[ 
  \mathcal{P}  (\sigma_z \otimes \mathbb{I}) 
 \Phi^n_q\right]\right|^2.
  \end{aligned}
\end{equation}
Here, $\Phi^n_q$, is shown in Eq.~(\ref{pwave}), denotes the excitation wave function of the excitation with energy $ \omega_n$ and quasimomentum $\hbar q$.  $n$ indexes the $n$-th band of the excitation spectrum, and  $\mathcal{P} = (\phi_1^*, \phi_2^*, \phi_1, \phi_2)$ representing the ground state, and $\mathbb{I}$ represents the $2\times2$ identity matrix. The symbol $\otimes$ denotes the tensor product (Kronecker product), which combines operators acting on different degrees of freedom: spin ($\sigma_z$) and particle-hole ($\mathbb{I}$). $S(q,\omega_n)$ and $ S_\sigma(q,\omega_n)$ represent density excitation and spin density excitation, respectively.

Figures~\ref{fig3}(b1) and \ref{fig3}(c1) show respectively density and spin-density structure factor corresponding to the excitation spectrum demonstrated in Fig.~\ref{fig3}(a1). The total factors of all excitation bands $S(q)=\sum_n S(q,\omega_n)$ and $S_\sigma(q)=\sum_n S_\sigma(q,\omega_n)$ are also shown by black lines in the figures. In the long-wavelength limit, the first band mainly belongs to density excitations [see the red line in Fig.~\ref{fig3}(b1)], while the second band belongs to spin-density excitations [see the blue line in Fig.~\ref{fig3}(c1)]. Around the energy avoided crossing between the first and second bands, $S(q,\omega_1)$ and $S(q,\omega_2)$, as well as $S_\sigma(q,\omega_1)$ and $S_\sigma(q,\omega_2)$, are swapping.  The first and second bands exchange characteristics: the first band becomes spin-density excitations and the second band takes on the nature of density excitations. In the region of $q/k_{\text{R}} \geqslant 1$, the factors of these two bands decrease, and the third band evolves into spin-density excitations [see green lines in Figs.~\ref{fig3}(b1) and \ref{fig3}(c1)].

The observed exchange of spin and density characteristics in the structure factor at the avoided crossing highlights an intriguing phenomenon. This spin exchange, along with the corresponding redistribution of excitation properties around the crossing, opens up a possibility for experimentally probing the avoided crossing structure with great precision. Such a potential measurement could provide valuable insights into the coupling mechanism  between density and spin-density modes and the transitions between different excitation regimes.

The energy avoided crossing between the lowest two excitation bands depends on the depth of the ZL $\Omega_\text{RF}$. As $\Omega_\text{RF}$ increases, the avoided crossing moves towards $q/k_{\text{R}} = 1$ and gradually disappears. Figure~\ref{fig3}(a2) presents its disappearance with parameters $(\Omega_{\text{R}}, \Omega_{\text{RF}})/E_{\text{R}} = (1, 1)$. The structure factors associated with the spectrum shown in Fig.~\ref{fig3}(a2) are illustrated in Figs.~\ref{fig3}(b2) and \ref{fig3}(c2). In the region of $ 0< q/k_{\text{R}} <1$, the first band corresponds to density excitations and the second band corresponds to spin-density excitations. In the region of $q/k_{\text{R}} >1$, similar to the case where the avoided crossing is present, the factors of the lowest two bands decrease to zero, but the third band evolves into density excitations [see the green lines in Figs.~\ref{fig3}(b2) and \ref{fig3}(c2)].

For comparison, we show the elementary excitations of a band-center state with parameters $ (\Omega_{\text{R}}, \Omega_{\text{RF}})/E_{\text{R}} = (6, 0.25)$ in Fig.~\ref{fig3}(a3). The excitation spectrum presents an obvious difference compared to the spectrum of a band-edge state in the geometry of the second and third bands. The structure factors described in Figs.~\ref{fig3}(b3) and \ref{fig3}(c3) demonstrate a dramatic difference compared to those of a band-edge state.  In the region of $ 0< q/k_{\text{R}} <1$, the first band is characterized by nonzero density and spin-density factors [see the red lines] and there is no contribution from the second band. Both factors are swapped around $q/k_{\text{R}} =1$, leading to the dominance of contributions from the second band in the region $q/k_{\text{R}} >1$ [see the blue lines]. It is interesting to note that the third band does not contribute to these two-type of excitations [see the green lines]. 

\begin{figure}[t]
	\centerline{\includegraphics[ width=0.5\textwidth]{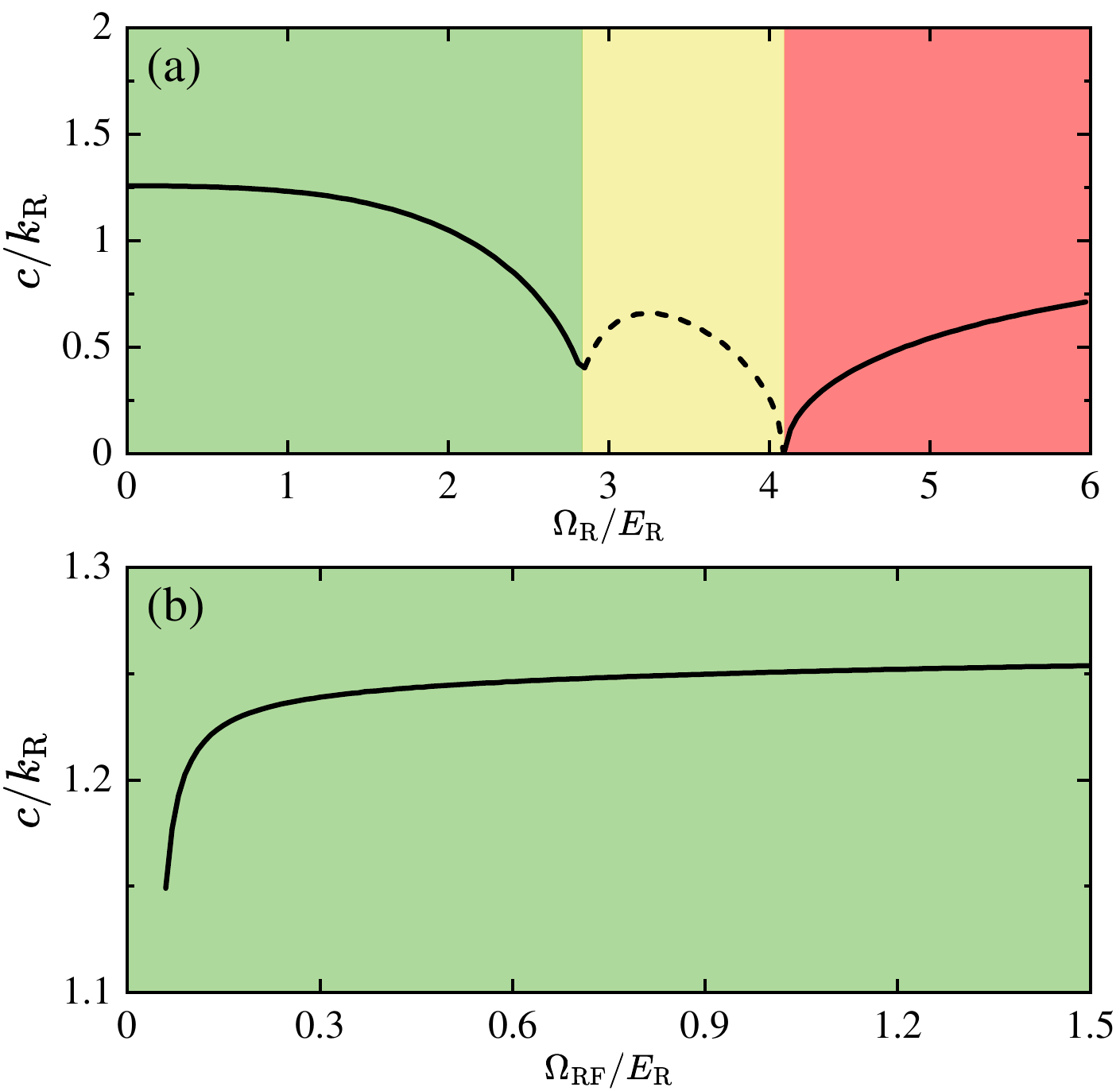}}
	\caption{ The speed of sound $c$  with the interaction coefficient $g k_\text{R}/E_\text{R} = 0.0848$. 
	(a) The speed of sound evolves across different phases. Regime of the band-edge (band-center) phase is represented by green (red) color filling. The yellow area denotes the single-momentum Bloch phase. The depth of  the ZL is fixed as \(\Omega_{\text{RF}}/E_{\text{R}} = 0.25\). (b) The speed of sound for the band-edge states as a function of the depth of the ZL with a  fixed  \(\Omega_{\text{R}}/E_{\text{R}} = 1\). 
	}
	\label{fig4}
\end{figure}

{The linear dispersion of the lowest excitation band in the long-wavelength limit shown in Fig.~\ref{fig3} is a signature of superfluidity of the ground states. The speed of sound $c$ is relevant to the slope of the linear dispersion. Based on the elementary excitations, we calculate the speed of sound.  Fig.~\ref{fig4}(a) demonstrates evolution of the speed of sound in different phases which are represented by different colors.  In different phases, the speed of sound behaves very differently. Around boundaries of phase transitions, the speed of sound is continuous, which indicates the phase transitions are second order.  In Fig.~\ref{fig4}(b), the speed of sound for the band-edge states is shown as a function of 
 $\Omega_\text{RF}$. In most of  $\Omega_\text{RF}$, the speed of sound almost keeps constant. In the small $\Omega_\text{RF}$ regime, the speed of sound decreases as $\Omega_\text{RF}$ decreases.  This is because that with a small $\Omega_\text{RF}$ the system trends towards to the phase with double energy minima from the band-edge states. }

\subsection{Superfluid density of the band-edge states}
\label{Superfluid}
Superfluid density closely relates to the speed of sound.  In the presence of optical lattices, the superfluid density of a ground-state BEC has been experimentally measured~\cite{PhysRevLett.130.226003,PhysRevLett.131.163401}. We calculate superfluid fraction of the band-edge states in the SOC ZL.  We first find the band-edge states of the GPE in Eq.~(\ref{nonGPE}) and calculate $p=\langle\Psi|(-i\hbar\partial_x) |  \Psi\rangle $. Then, imagine that there is a perturbation potential moving at a velocity $v$. In the co-moving reference frame, the superfluid exhibits a relative counter-flow with respect to the imaginary potential, and the normal fluid becomes stationary due to it moving together with the potential~\cite{leggett2006quantum}. 
The GPE in Eq.~(\ref{nonGPE}) in the co-moving frame with velocity $v$ becomes
\begin{equation}
\label{movingGPE}
    i\hbar \frac{\partial  \Psi^\prime}{\partial t} = \left[ H_\text{sin} + H_\text{non}  +i\hbar v\partial_x  \right] \Psi^\prime,
\end{equation}
where $\Psi^\prime$ represents the wave function in the co-moving frame. 
For a given $v$, we find the band-edge states of Eq.~(\ref{movingGPE}) and calculate $p^\prime=\langle\Psi^\prime|(-i\hbar\partial_x) |  \Psi^\prime\rangle $. Based on the difference of $p$ and $p'$, the superfluid fraction of the band-edge states is defined as~\cite{PhysRevLett.130.226003,PhysRevLett.131.163401,biagioni2024measurement}
\begin{equation}
f_s = 1 - \left| \lim_{v \to 0}\frac{p^\prime-p }{m v} \right|.
\end{equation}


\begin{figure}[t]
	\centerline{\includegraphics[ width=0.5\textwidth]{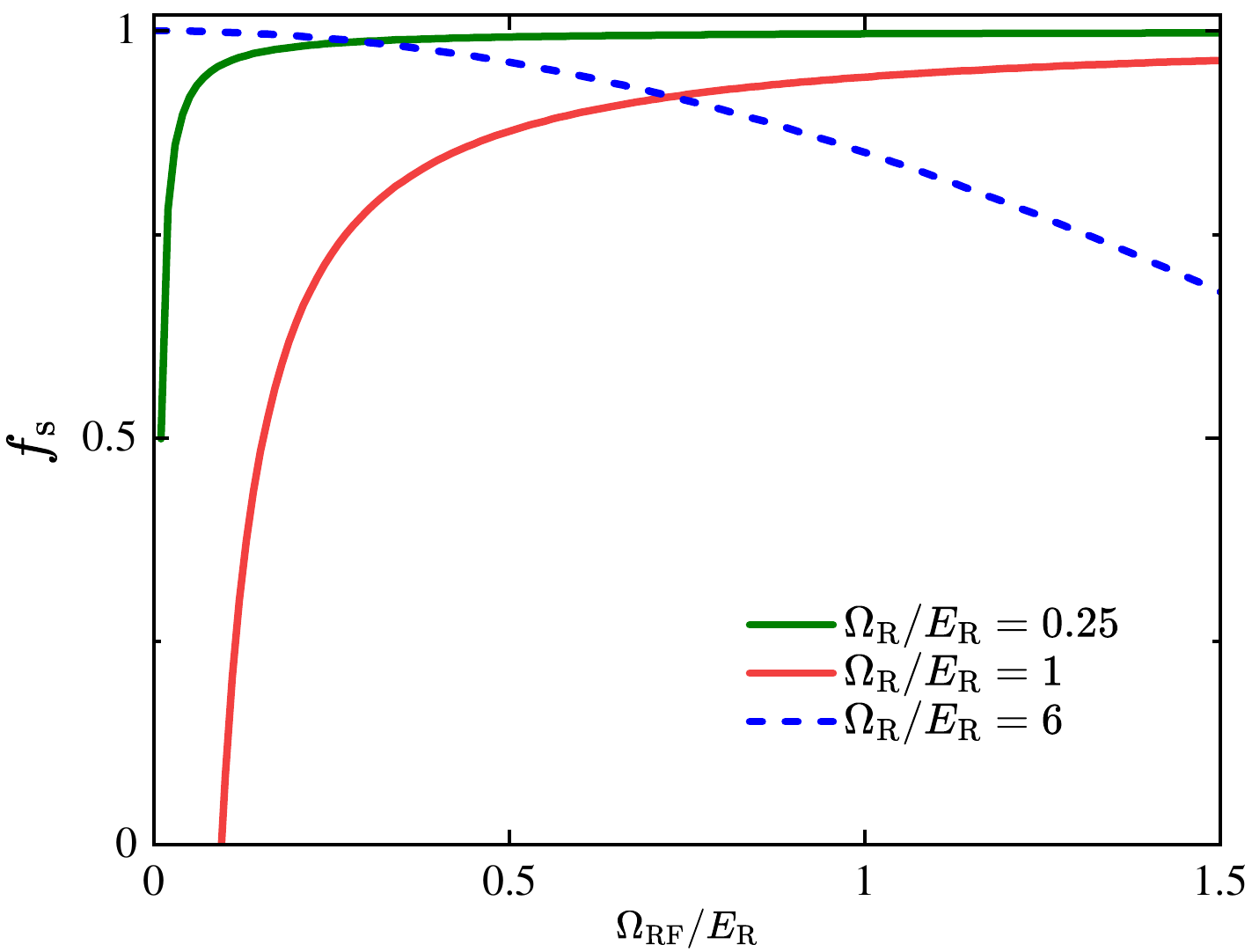}}
	\caption{
		Superfluid fraction $f_s$ of the band-edge states (green and red solid lines) and the band-center states (blue-dashed line) as a function of the depth of the ZL $\Omega_{\text{RF}}$, for the parameters $\Omega_{\text{R}}/E_{\text{R}} = 0.25$ (green solid line), $\Omega_{\text{R}}/E_{\text{R}} = 1$ (red solid line), and $\Omega_{\text{R}}/E_{\text{R}} = 6$ (blue dashed line). }
	\label{fig5}
\end{figure}

The calculated superfluid fraction of the band-edge states is demonstrated in Fig.~\ref{fig5}. For the band-edge states with $\Omega_{\text{R}}/E_{\text{R}} = 0.25$, the superfluid fraction is almost equal to 1 [see the green line].  The red line shows the result for the band-edge states with $\Omega_{\text{R}}/E_{\text{R}} = 1$. As the lattice depth $\Omega_{\text{RF}}$ decreases, the superfluid fraction decreases dramatically. For a larger $\Omega_{\text{RF}}$, it continuously converges towards $1$. The dramatic decrease of the superfluid fraction with decreasing $\Omega_{\text{RF}}$ is due to the fact that the system is preparing for the transition between the band-edge phase and the phase with double energy minima. We emphasize that the superfluid fraction of the band-edge states has a large value when $\Omega_{\text{RF}}$ is large. However, {it is interesting to point out that the spin-orbit coupling plays an important role. Only with the spin-orbit coupling, the band-edge states may have superfluidity.}  It is the spin-orbit coupling that makes the band-edge phase becomes ground state. Without the spin-orbit coupling, we can also find the existence of Bloch states in the lowest band at the Brillouin zone edges of optical lattices or the ZL. However, superfluidity of these Bloch states is broken down by the dynamical and energetic instabilities, so there is no superfluid fraction involved in these band-edge Bloch states~\cite{PhysRevA.64.061603}.

For comparison, we also calculate the superfluid fraction for the band-center states with $\Omega_{\text{R}}/E_{\text{R}} = 6$, and the result is shown as the blue-dashed line in Fig.~\ref{fig5}. The superfluid fraction decreases as $\Omega_{\text{RF}}$ increases.
 This is because the effective mass of the band-center states increases as the lattice depth increases. The superfluid fraction is inversely proportional to the effective mass~\cite{PhysRevLett.130.226003,PhysRevLett.131.163401}.

\section{Conclusion}
\label{conclusion}

The systems of BECs in the SOC ZL have an important role in ultracold atomic physics. We systematically study their ground-state phase diagram. The SOC ZL features the Bloch band-gap spectrum. Depending on the location and number of energy minima in the lowest Bloch band, the single-particle phase diagram is identified in the parameter space of the Raman coupling $\Omega_\text{R}$ and the depth of the ZL $\Omega_\text{RF}$ and is classified into three phases: the band-edge phase, with energy minima of the lowest Bloch band located at the Brillouin zone edges; the band-center phase, with the energy minimum of the lowest Bloch band located at the Brillouin zone center; and the phase with two energy minima located inside the first Brillouin zone. How interacting atoms condense into these phases constitutes the ground-state phase diagram. In the band-edge and band-center phases, there is only one choice (i.e., the band-edge Bloch state or the band-center Bloch state) for atoms to condense. In the third phase, atoms may condense into one of the energy minima, forming a single-momentum Bloch state, or may simultaneously occupy two energy minima, forming a stripe state.

One of the outstanding features of BECs in the SOC ZL is that the band-edge phase exists in a very broad parameter region. The band-edge states are {unique} and do not have an analogue in SOC homogeneous systems or in lattice systems without spin-orbit coupling. We identify the superfluidity of the band-edge states by calculating elementary excitations and the superfluid fraction.  {There is an energy avoided crossing between the gapless mode and gaped pseudo-Goldstone mode in the elementary excitations, which can be experimentally observed by Bragg spectroscopy.}

\begin{acknowledgments}
This work is supported by the National Natural Science Foundation of China (NSFC) under Grants No. 12374247 and No. 11974235, as well as by the Shanghai Municipal Science and Technology Major Project (Grant No. 2019SHZDZX01-ZX04). 

\end{acknowledgments}

\bibliography{ref}

\begin{thebibliography}{89}%
\makeatletter
\providecommand \@ifxundefined [1]{%
 \@ifx{#1\undefined}
}%
\providecommand \@ifnum [1]{%
 \ifnum #1\expandafter \@firstoftwo
 \else \expandafter \@secondoftwo
 \fi
}%
\providecommand \@ifx [1]{%
 \ifx #1\expandafter \@firstoftwo
 \else \expandafter \@secondoftwo
 \fi
}%
\providecommand \natexlab [1]{#1}%
\providecommand \enquote  [1]{``#1''}%
\providecommand \bibnamefont  [1]{#1}%
\providecommand \bibfnamefont [1]{#1}%
\providecommand \citenamefont [1]{#1}%
\providecommand \href@noop [0]{\@secondoftwo}%
\providecommand \href [0]{\begingroup \@sanitize@url \@href}%
\providecommand \@href[1]{\@@startlink{#1}\@@href}%
\providecommand \@@href[1]{\endgroup#1\@@endlink}%
\providecommand \@sanitize@url [0]{\catcode `\\12\catcode `\$12\catcode
  `\&12\catcode `\#12\catcode `\^12\catcode `\_12\catcode `\%12\relax}%
\providecommand \@@startlink[1]{}%
\providecommand \@@endlink[0]{}%
\providecommand \url  [0]{\begingroup\@sanitize@url \@url }%
\providecommand \@url [1]{\endgroup\@href {#1}{\urlprefix }}%
\providecommand \urlprefix  [0]{URL }%
\providecommand \Eprint [0]{\href }%
\providecommand \doibase [0]{https://doi.org/}%
\providecommand \selectlanguage [0]{\@gobble}%
\providecommand \bibinfo  [0]{\@secondoftwo}%
\providecommand \bibfield  [0]{\@secondoftwo}%
\providecommand \translation [1]{[#1]}%
\providecommand \BibitemOpen [0]{}%
\providecommand \bibitemStop [0]{}%
\providecommand \bibitemNoStop [0]{.\EOS\space}%
\providecommand \EOS [0]{\spacefactor3000\relax}%
\providecommand \BibitemShut  [1]{\csname bibitem#1\endcsname}%
\let\auto@bib@innerbib\@empty
\bibitem [{\citenamefont {Lin}\ \emph {et~al.}(2011)\citenamefont {Lin},
  \citenamefont {Jim{\'e}nez-Garc{\'\i}a},\ and\ \citenamefont
  {Spielman}}]{lin2011spin}%
  \BibitemOpen
  \bibfield  {author} {\bibinfo {author} {\bibfnamefont {Y.-J.}\ \bibnamefont
  {Lin}}, \bibinfo {author} {\bibfnamefont {K.}~\bibnamefont
  {Jim{\'e}nez-Garc{\'\i}a}},\ and\ \bibinfo {author} {\bibfnamefont {I.~B.}\
  \bibnamefont {Spielman}},\ }\bibfield  {title} {\bibinfo {title}
  {{Spin--orbit-coupled Bose--Einstein condensates}},\ }\href
  {https://www.nature.com/articles/nature09887} {\bibfield  {journal} {\bibinfo
   {journal} {Nature}\ }\textbf {\bibinfo {volume} {471}},\ \bibinfo {pages}
  {83} (\bibinfo {year} {2011})}\BibitemShut {NoStop}%
\bibitem [{\citenamefont {Huang}\ \emph {et~al.}(2016)\citenamefont {Huang},
  \citenamefont {Meng}, \citenamefont {Wang}, \citenamefont {Peng},
  \citenamefont {Zhang}, \citenamefont {Chen}, \citenamefont {Li},
  \citenamefont {Zhou},\ and\ \citenamefont {Zhang}}]{huang2016experimental}%
  \BibitemOpen
  \bibfield  {author} {\bibinfo {author} {\bibfnamefont {L.}~\bibnamefont
  {Huang}}, \bibinfo {author} {\bibfnamefont {Z.}~\bibnamefont {Meng}},
  \bibinfo {author} {\bibfnamefont {P.}~\bibnamefont {Wang}}, \bibinfo {author}
  {\bibfnamefont {P.}~\bibnamefont {Peng}}, \bibinfo {author} {\bibfnamefont
  {S.-L.}\ \bibnamefont {Zhang}}, \bibinfo {author} {\bibfnamefont
  {L.}~\bibnamefont {Chen}}, \bibinfo {author} {\bibfnamefont {D.}~\bibnamefont
  {Li}}, \bibinfo {author} {\bibfnamefont {Q.}~\bibnamefont {Zhou}},\ and\
  \bibinfo {author} {\bibfnamefont {J.}~\bibnamefont {Zhang}},\ }\bibfield
  {title} {\bibinfo {title} {{Experimental realization of two-dimensional
  synthetic spin--orbit coupling in ultracold Fermi gases}},\ }\href
  {https://doi.org/10.1038/nphys3672} {\bibfield  {journal} {\bibinfo
  {journal} {Nat. Phys.}\ }\textbf {\bibinfo {volume} {12}},\ \bibinfo {pages}
  {540} (\bibinfo {year} {2016})}\BibitemShut {NoStop}%
\bibitem [{\citenamefont {Wu}\ \emph {et~al.}(2016)\citenamefont {Wu},
  \citenamefont {Zhang}, \citenamefont {Sun}, \citenamefont {Xu}, \citenamefont
  {Wang}, \citenamefont {Ji}, \citenamefont {Deng}, \citenamefont {Chen},
  \citenamefont {Liu},\ and\ \citenamefont {Pan}}]{wu2016realization}%
  \BibitemOpen
  \bibfield  {author} {\bibinfo {author} {\bibfnamefont {Z.}~\bibnamefont
  {Wu}}, \bibinfo {author} {\bibfnamefont {L.}~\bibnamefont {Zhang}}, \bibinfo
  {author} {\bibfnamefont {W.}~\bibnamefont {Sun}}, \bibinfo {author}
  {\bibfnamefont {X.-T.}\ \bibnamefont {Xu}}, \bibinfo {author} {\bibfnamefont
  {B.-Z.}\ \bibnamefont {Wang}}, \bibinfo {author} {\bibfnamefont {S.-C.}\
  \bibnamefont {Ji}}, \bibinfo {author} {\bibfnamefont {Y.}~\bibnamefont
  {Deng}}, \bibinfo {author} {\bibfnamefont {S.}~\bibnamefont {Chen}}, \bibinfo
  {author} {\bibfnamefont {X.-J.}\ \bibnamefont {Liu}},\ and\ \bibinfo {author}
  {\bibfnamefont {J.-W.}\ \bibnamefont {Pan}},\ }\bibfield  {title} {\bibinfo
  {title} {{Realization of two-dimensional spin-orbit coupling for
  Bose-Einstein condensates}},\ }\href
  {https://www.science.org/doi/full/10.1126/science.aaf6689?casa_token=emS1u22KuhEAAAAA%3A45Bcds2JViYxnz8NbPSK8IAmoACEnmcULutBKXu1nF6OzclGN_XB5yNJXUtuEZGfPBB9wHqQEj-1Jg}
  {\bibfield  {journal} {\bibinfo  {journal} {Science}\ }\textbf {\bibinfo
  {volume} {354}},\ \bibinfo {pages} {83} (\bibinfo {year} {2016})}\BibitemShut
  {NoStop}%
\bibitem [{\citenamefont {Goldman}\ \emph {et~al.}(2014)\citenamefont
  {Goldman}, \citenamefont {Juzeli{\=u}nas}, \citenamefont {{\"O}hberg},\ and\
  \citenamefont {Spielman}}]{goldman2014light}%
  \BibitemOpen
  \bibfield  {author} {\bibinfo {author} {\bibfnamefont {N.}~\bibnamefont
  {Goldman}}, \bibinfo {author} {\bibfnamefont {G.}~\bibnamefont
  {Juzeli{\=u}nas}}, \bibinfo {author} {\bibfnamefont {P.}~\bibnamefont
  {{\"O}hberg}},\ and\ \bibinfo {author} {\bibfnamefont {I.~B.}\ \bibnamefont
  {Spielman}},\ }\bibfield  {title} {\bibinfo {title} {Light-induced gauge
  fields for ultracold atoms},\ }\href
  {https://iopscience.iop.org/article/10.1088/0034-4885/77/12/126401/meta}
  {\bibfield  {journal} {\bibinfo  {journal} {Rep. Prog. Phys.}\ }\textbf
  {\bibinfo {volume} {77}},\ \bibinfo {pages} {126401} (\bibinfo {year}
  {2014})}\BibitemShut {NoStop}%
\bibitem [{\citenamefont {Zhai}(2015)}]{zhai2015degenerate}%
  \BibitemOpen
  \bibfield  {author} {\bibinfo {author} {\bibfnamefont {H.}~\bibnamefont
  {Zhai}},\ }\bibfield  {title} {\bibinfo {title} {{Degenerate quantum gases
  with spin--orbit coupling: a review}},\ }\href
  {https://iopscience.iop.org/article/10.1088/0034-4885/78/2/026001/meta?casa_token=1UELOIHg1KEAAAAA:M3wUCA78udT-VqnuCxVMIl0hr9x2fdypRikwsCMyG4pPEY3EljYkYdZFDVOsHsIZ1CAfF_GaFCNvW0R-z5HgPn31rHM}
  {\bibfield  {journal} {\bibinfo  {journal} {Rep. Prog. Phys.}\ }\textbf
  {\bibinfo {volume} {78}},\ \bibinfo {pages} {026001} (\bibinfo {year}
  {2015})}\BibitemShut {NoStop}%
\bibitem [{\citenamefont {Zhang}\ \emph
  {et~al.}(2016{\natexlab{a}})\citenamefont {Zhang}, \citenamefont {Mossman},
  \citenamefont {Busch}, \citenamefont {Engels},\ and\ \citenamefont
  {Zhang}}]{zhang2016properties}%
  \BibitemOpen
  \bibfield  {author} {\bibinfo {author} {\bibfnamefont {Y.}~\bibnamefont
  {Zhang}}, \bibinfo {author} {\bibfnamefont {M.~E.}\ \bibnamefont {Mossman}},
  \bibinfo {author} {\bibfnamefont {T.}~\bibnamefont {Busch}}, \bibinfo
  {author} {\bibfnamefont {P.}~\bibnamefont {Engels}},\ and\ \bibinfo {author}
  {\bibfnamefont {C.}~\bibnamefont {Zhang}},\ }\bibfield  {title} {\bibinfo
  {title} {{Properties of spin-orbit-coupled Bose-Einstein condensates}},\
  }\href {https://link.springer.com/article/10.1007/s11467-016-0560-y}
  {\bibfield  {journal} {\bibinfo  {journal} {Front. Phys.}\ }\textbf {\bibinfo
  {volume} {11}},\ \bibinfo {pages} {1} (\bibinfo {year}
  {2016}{\natexlab{a}})}\BibitemShut {NoStop}%
\bibitem [{\citenamefont {Wang}\ \emph {et~al.}(2010)\citenamefont {Wang},
  \citenamefont {Gao}, \citenamefont {Jian},\ and\ \citenamefont
  {Zhai}}]{PhysRevLett.105.160403}%
  \BibitemOpen
  \bibfield  {author} {\bibinfo {author} {\bibfnamefont {C.}~\bibnamefont
  {Wang}}, \bibinfo {author} {\bibfnamefont {C.}~\bibnamefont {Gao}}, \bibinfo
  {author} {\bibfnamefont {C.-M.}\ \bibnamefont {Jian}},\ and\ \bibinfo
  {author} {\bibfnamefont {H.}~\bibnamefont {Zhai}},\ }\bibfield  {title}
  {\bibinfo {title} {{Spin-Orbit Coupled Spinor Bose-Einstein Condensates}},\
  }\href {https://doi.org/10.1103/PhysRevLett.105.160403} {\bibfield  {journal}
  {\bibinfo  {journal} {Phys. Rev. Lett.}\ }\textbf {\bibinfo {volume} {105}},\
  \bibinfo {pages} {160403} (\bibinfo {year} {2010})}\BibitemShut {NoStop}%
\bibitem [{\citenamefont {Ho}\ and\ \citenamefont
  {Zhang}(2011)}]{PhysRevLett.107.150403}%
  \BibitemOpen
  \bibfield  {author} {\bibinfo {author} {\bibfnamefont {T.-L.}\ \bibnamefont
  {Ho}}\ and\ \bibinfo {author} {\bibfnamefont {S.}~\bibnamefont {Zhang}},\
  }\bibfield  {title} {\bibinfo {title} {{Bose-Einstein Condensates with
  Spin-Orbit Interaction}},\ }\href
  {https://doi.org/10.1103/PhysRevLett.107.150403} {\bibfield  {journal}
  {\bibinfo  {journal} {Phys. Rev. Lett.}\ }\textbf {\bibinfo {volume} {107}},\
  \bibinfo {pages} {150403} (\bibinfo {year} {2011})}\BibitemShut {NoStop}%
\bibitem [{\citenamefont {Hu}\ \emph {et~al.}(2012)\citenamefont {Hu},
  \citenamefont {Ramachandhran}, \citenamefont {Pu},\ and\ \citenamefont
  {Liu}}]{PhysRevLett.108.010402}%
  \BibitemOpen
  \bibfield  {author} {\bibinfo {author} {\bibfnamefont {H.}~\bibnamefont
  {Hu}}, \bibinfo {author} {\bibfnamefont {B.}~\bibnamefont {Ramachandhran}},
  \bibinfo {author} {\bibfnamefont {H.}~\bibnamefont {Pu}},\ and\ \bibinfo
  {author} {\bibfnamefont {X.-J.}\ \bibnamefont {Liu}},\ }\bibfield  {title}
  {\bibinfo {title} {{Spin-Orbit Coupled Weakly Interacting Bose-Einstein
  Condensates in Harmonic Traps}},\ }\href
  {https://doi.org/10.1103/PhysRevLett.108.010402} {\bibfield  {journal}
  {\bibinfo  {journal} {Phys. Rev. Lett.}\ }\textbf {\bibinfo {volume} {108}},\
  \bibinfo {pages} {010402} (\bibinfo {year} {2012})}\BibitemShut {NoStop}%
\bibitem [{\citenamefont {Li}\ \emph {et~al.}(2012)\citenamefont {Li},
  \citenamefont {Pitaevskii},\ and\ \citenamefont
  {Stringari}}]{PhysRevLett.108.225301}%
  \BibitemOpen
  \bibfield  {author} {\bibinfo {author} {\bibfnamefont {Y.}~\bibnamefont
  {Li}}, \bibinfo {author} {\bibfnamefont {L.~P.}\ \bibnamefont {Pitaevskii}},\
  and\ \bibinfo {author} {\bibfnamefont {S.}~\bibnamefont {Stringari}},\
  }\bibfield  {title} {\bibinfo {title} {{Quantum Tricriticality and Phase
  Transitions in Spin-Orbit Coupled Bose-Einstein Condensates}},\ }\href
  {https://doi.org/10.1103/PhysRevLett.108.225301} {\bibfield  {journal}
  {\bibinfo  {journal} {Phys. Rev. Lett.}\ }\textbf {\bibinfo {volume} {108}},\
  \bibinfo {pages} {225301} (\bibinfo {year} {2012})}\BibitemShut {NoStop}%
\bibitem [{\citenamefont {Yu}(2016)}]{PhysRevA.93.033648}%
  \BibitemOpen
  \bibfield  {author} {\bibinfo {author} {\bibfnamefont {Z.-Q.}\ \bibnamefont
  {Yu}},\ }\bibfield  {title} {\bibinfo {title} {(phase transitions and
  elementary excitations in spin-1 bose gases with raman-induced spin-orbit
  coupling)},\ }\href {https://doi.org/10.1103/PhysRevA.93.033648} {\bibfield
  {journal} {\bibinfo  {journal} {Phys. Rev. A}\ }\textbf {\bibinfo {volume}
  {93}},\ \bibinfo {pages} {033648} (\bibinfo {year} {2016})}\BibitemShut
  {NoStop}%
\bibitem [{\citenamefont {Hamner}\ \emph {et~al.}(2014)\citenamefont {Hamner},
  \citenamefont {Qu}, \citenamefont {Zhang}, \citenamefont {Chang},
  \citenamefont {Gong}, \citenamefont {Zhang},\ and\ \citenamefont
  {Engels}}]{hamner2014dicke}%
  \BibitemOpen
  \bibfield  {author} {\bibinfo {author} {\bibfnamefont {C.}~\bibnamefont
  {Hamner}}, \bibinfo {author} {\bibfnamefont {C.}~\bibnamefont {Qu}}, \bibinfo
  {author} {\bibfnamefont {Y.}~\bibnamefont {Zhang}}, \bibinfo {author}
  {\bibfnamefont {J.}~\bibnamefont {Chang}}, \bibinfo {author} {\bibfnamefont
  {M.}~\bibnamefont {Gong}}, \bibinfo {author} {\bibfnamefont {C.}~\bibnamefont
  {Zhang}},\ and\ \bibinfo {author} {\bibfnamefont {P.}~\bibnamefont
  {Engels}},\ }\bibfield  {title} {\bibinfo {title} {{Dicke-type phase
  transition in a spin-orbit-coupled Bose-Einstein condensate}},\ }\href
  {https://www.nature.com/articles/ncomms5023} {\bibfield  {journal} {\bibinfo
  {journal} {Nat. Commun.}\ }\textbf {\bibinfo {volume} {5}},\ \bibinfo {pages}
  {4023} (\bibinfo {year} {2014})}\BibitemShut {NoStop}%
\bibitem [{\citenamefont {Ji}\ \emph {et~al.}(2014)\citenamefont {Ji},
  \citenamefont {Zhang}, \citenamefont {Zhang}, \citenamefont {Du},
  \citenamefont {Zheng}, \citenamefont {Deng}, \citenamefont {Zhai},
  \citenamefont {Chen},\ and\ \citenamefont {Pan}}]{ji2014experimental}%
  \BibitemOpen
  \bibfield  {author} {\bibinfo {author} {\bibfnamefont {S.-C.}\ \bibnamefont
  {Ji}}, \bibinfo {author} {\bibfnamefont {J.-Y.}\ \bibnamefont {Zhang}},
  \bibinfo {author} {\bibfnamefont {L.}~\bibnamefont {Zhang}}, \bibinfo
  {author} {\bibfnamefont {Z.-D.}\ \bibnamefont {Du}}, \bibinfo {author}
  {\bibfnamefont {W.}~\bibnamefont {Zheng}}, \bibinfo {author} {\bibfnamefont
  {Y.-J.}\ \bibnamefont {Deng}}, \bibinfo {author} {\bibfnamefont
  {H.}~\bibnamefont {Zhai}}, \bibinfo {author} {\bibfnamefont {S.}~\bibnamefont
  {Chen}},\ and\ \bibinfo {author} {\bibfnamefont {J.-W.}\ \bibnamefont
  {Pan}},\ }\bibfield  {title} {\bibinfo {title} {{Experimental determination
  of the finite-temperature phase diagram of a spin--orbit coupled Bose gas}},\
  }\href {https://www.nature.com/articles/nphys2905} {\bibfield  {journal}
  {\bibinfo  {journal} {Nat. Phys.}\ }\textbf {\bibinfo {volume} {10}},\
  \bibinfo {pages} {314} (\bibinfo {year} {2014})}\BibitemShut {NoStop}%
\bibitem [{\citenamefont {Jim\'enez-Garc\'{\i}a}\ \emph
  {et~al.}(2015)\citenamefont {Jim\'enez-Garc\'{\i}a}, \citenamefont {LeBlanc},
  \citenamefont {Williams}, \citenamefont {Beeler}, \citenamefont {Qu},
  \citenamefont {Gong}, \citenamefont {Zhang},\ and\ \citenamefont
  {Spielman}}]{PhysRevLett.114.125301}%
  \BibitemOpen
  \bibfield  {author} {\bibinfo {author} {\bibfnamefont {K.}~\bibnamefont
  {Jim\'enez-Garc\'{\i}a}}, \bibinfo {author} {\bibfnamefont {L.~J.}\
  \bibnamefont {LeBlanc}}, \bibinfo {author} {\bibfnamefont {R.~A.}\
  \bibnamefont {Williams}}, \bibinfo {author} {\bibfnamefont {M.~C.}\
  \bibnamefont {Beeler}}, \bibinfo {author} {\bibfnamefont {C.}~\bibnamefont
  {Qu}}, \bibinfo {author} {\bibfnamefont {M.}~\bibnamefont {Gong}}, \bibinfo
  {author} {\bibfnamefont {C.}~\bibnamefont {Zhang}},\ and\ \bibinfo {author}
  {\bibfnamefont {I.~B.}\ \bibnamefont {Spielman}},\ }\bibfield  {title}
  {\bibinfo {title} {{Tunable Spin-Orbit Coupling via Strong Driving in
  Ultracold-Atom Systems}},\ }\href
  {https://doi.org/10.1103/PhysRevLett.114.125301} {\bibfield  {journal}
  {\bibinfo  {journal} {Phys. Rev. Lett.}\ }\textbf {\bibinfo {volume} {114}},\
  \bibinfo {pages} {125301} (\bibinfo {year} {2015})}\BibitemShut {NoStop}%
\bibitem [{\citenamefont {Li}\ \emph {et~al.}(2017)\citenamefont {Li},
  \citenamefont {Lee}, \citenamefont {Huang}, \citenamefont {Burchesky},
  \citenamefont {Shteynas}, \citenamefont {Top}, \citenamefont {Jamison},\ and\
  \citenamefont {Ketterle}}]{li2017stripe}%
  \BibitemOpen
  \bibfield  {author} {\bibinfo {author} {\bibfnamefont {J.-R.}\ \bibnamefont
  {Li}}, \bibinfo {author} {\bibfnamefont {J.}~\bibnamefont {Lee}}, \bibinfo
  {author} {\bibfnamefont {W.}~\bibnamefont {Huang}}, \bibinfo {author}
  {\bibfnamefont {S.}~\bibnamefont {Burchesky}}, \bibinfo {author}
  {\bibfnamefont {B.}~\bibnamefont {Shteynas}}, \bibinfo {author}
  {\bibfnamefont {F.~{\c{C}}.}\ \bibnamefont {Top}}, \bibinfo {author}
  {\bibfnamefont {A.~O.}\ \bibnamefont {Jamison}},\ and\ \bibinfo {author}
  {\bibfnamefont {W.}~\bibnamefont {Ketterle}},\ }\bibfield  {title} {\bibinfo
  {title} {{A stripe phase with supersolid properties in spin--orbit-coupled
  Bose--Einstein condensates}},\ }\href
  {https://www.nature.com/articles/nature21431} {\bibfield  {journal} {\bibinfo
   {journal} {Nature}\ }\textbf {\bibinfo {volume} {543}},\ \bibinfo {pages}
  {91} (\bibinfo {year} {2017})}\BibitemShut {NoStop}%
\bibitem [{\citenamefont {Martone}\ \emph {et~al.}(2012)\citenamefont
  {Martone}, \citenamefont {Li}, \citenamefont {Pitaevskii},\ and\
  \citenamefont {Stringari}}]{PhysRevA.86.063621}%
  \BibitemOpen
  \bibfield  {author} {\bibinfo {author} {\bibfnamefont {G.~I.}\ \bibnamefont
  {Martone}}, \bibinfo {author} {\bibfnamefont {Y.}~\bibnamefont {Li}},
  \bibinfo {author} {\bibfnamefont {L.~P.}\ \bibnamefont {Pitaevskii}},\ and\
  \bibinfo {author} {\bibfnamefont {S.}~\bibnamefont {Stringari}},\ }\bibfield
  {title} {\bibinfo {title} {{Anisotropic dynamics of a spin-orbit-coupled
  Bose-Einstein condensate}},\ }\href
  {https://doi.org/10.1103/PhysRevA.86.063621} {\bibfield  {journal} {\bibinfo
  {journal} {Phys. Rev. A}\ }\textbf {\bibinfo {volume} {86}},\ \bibinfo
  {pages} {063621} (\bibinfo {year} {2012})}\BibitemShut {NoStop}%
\bibitem [{\citenamefont {Li}\ \emph {et~al.}(2013)\citenamefont {Li},
  \citenamefont {Martone}, \citenamefont {Pitaevskii},\ and\ \citenamefont
  {Stringari}}]{PhysRevLett.110.235302}%
  \BibitemOpen
  \bibfield  {author} {\bibinfo {author} {\bibfnamefont {Y.}~\bibnamefont
  {Li}}, \bibinfo {author} {\bibfnamefont {G.~I.}\ \bibnamefont {Martone}},
  \bibinfo {author} {\bibfnamefont {L.~P.}\ \bibnamefont {Pitaevskii}},\ and\
  \bibinfo {author} {\bibfnamefont {S.}~\bibnamefont {Stringari}},\ }\bibfield
  {title} {\bibinfo {title} {{Superstripes and the Excitation Spectrum of a
  Spin-Orbit-Coupled Bose-Einstein Condensate}},\ }\href
  {https://doi.org/10.1103/PhysRevLett.110.235302} {\bibfield  {journal}
  {\bibinfo  {journal} {Phys. Rev. Lett.}\ }\textbf {\bibinfo {volume} {110}},\
  \bibinfo {pages} {235302} (\bibinfo {year} {2013})}\BibitemShut {NoStop}%
\bibitem [{\citenamefont {Zheng}\ \emph {et~al.}(2013)\citenamefont {Zheng},
  \citenamefont {Yu}, \citenamefont {Cui},\ and\ \citenamefont
  {Zhai}}]{zheng2013properties}%
  \BibitemOpen
  \bibfield  {author} {\bibinfo {author} {\bibfnamefont {W.}~\bibnamefont
  {Zheng}}, \bibinfo {author} {\bibfnamefont {Z.-Q.}\ \bibnamefont {Yu}},
  \bibinfo {author} {\bibfnamefont {X.}~\bibnamefont {Cui}},\ and\ \bibinfo
  {author} {\bibfnamefont {H.}~\bibnamefont {Zhai}},\ }\bibfield  {title}
  {\bibinfo {title} {{Properties of Bose gases with the Raman-induced
  spin-orbit coupling}},\ }\href
  {https://iopscience.iop.org/article/10.1088/0953-4075/46/13/134007/meta?casa_token=0dpMEpkz9gMAAAAA:4NMs6S_I7sRsEzkTndDvlUUouA_XtAC1EIYIBanjZB4Ps1Skj4ZcvL-yfD4gRNYPCF7Qk6gleBPUbXRd_6ynKpuYHm0}
  {\bibfield  {journal} {\bibinfo  {journal} {J. Phys. B: At., Mol. Opt. Phys}\
  }\textbf {\bibinfo {volume} {46}},\ \bibinfo {pages} {134007} (\bibinfo
  {year} {2013})}\BibitemShut {NoStop}%
\bibitem [{\citenamefont {Khamehchi}\ \emph {et~al.}(2014)\citenamefont
  {Khamehchi}, \citenamefont {Zhang}, \citenamefont {Hamner}, \citenamefont
  {Busch},\ and\ \citenamefont {Engels}}]{PhysRevA.90.063624}%
  \BibitemOpen
  \bibfield  {author} {\bibinfo {author} {\bibfnamefont {M.~A.}\ \bibnamefont
  {Khamehchi}}, \bibinfo {author} {\bibfnamefont {Y.}~\bibnamefont {Zhang}},
  \bibinfo {author} {\bibfnamefont {C.}~\bibnamefont {Hamner}}, \bibinfo
  {author} {\bibfnamefont {T.}~\bibnamefont {Busch}},\ and\ \bibinfo {author}
  {\bibfnamefont {P.}~\bibnamefont {Engels}},\ }\bibfield  {title} {\bibinfo
  {title} {{Measurement of collective excitations in a spin-orbit-coupled
  Bose-Einstein condensate}},\ }\href
  {https://doi.org/10.1103/PhysRevA.90.063624} {\bibfield  {journal} {\bibinfo
  {journal} {Phys. Rev. A}\ }\textbf {\bibinfo {volume} {90}},\ \bibinfo
  {pages} {063624} (\bibinfo {year} {2014})}\BibitemShut {NoStop}%
\bibitem [{\citenamefont {Ji}\ \emph {et~al.}(2015)\citenamefont {Ji},
  \citenamefont {Zhang}, \citenamefont {Xu}, \citenamefont {Wu}, \citenamefont
  {Deng}, \citenamefont {Chen},\ and\ \citenamefont
  {Pan}}]{PhysRevLett.114.105301}%
  \BibitemOpen
  \bibfield  {author} {\bibinfo {author} {\bibfnamefont {S.-C.}\ \bibnamefont
  {Ji}}, \bibinfo {author} {\bibfnamefont {L.}~\bibnamefont {Zhang}}, \bibinfo
  {author} {\bibfnamefont {X.-T.}\ \bibnamefont {Xu}}, \bibinfo {author}
  {\bibfnamefont {Z.}~\bibnamefont {Wu}}, \bibinfo {author} {\bibfnamefont
  {Y.}~\bibnamefont {Deng}}, \bibinfo {author} {\bibfnamefont {S.}~\bibnamefont
  {Chen}},\ and\ \bibinfo {author} {\bibfnamefont {J.-W.}\ \bibnamefont
  {Pan}},\ }\bibfield  {title} {\bibinfo {title} {{Softening of Roton and
  Phonon Modes in a Bose-Einstein Condensate with Spin-Orbit Coupling}},\
  }\href {https://doi.org/10.1103/PhysRevLett.114.105301} {\bibfield  {journal}
  {\bibinfo  {journal} {Phys. Rev. Lett.}\ }\textbf {\bibinfo {volume} {114}},\
  \bibinfo {pages} {105301} (\bibinfo {year} {2015})}\BibitemShut {NoStop}%
\bibitem [{\citenamefont {Lyu}\ and\ \citenamefont
  {Zhang}(2020)}]{PhysRevA.102.023327}%
  \BibitemOpen
  \bibfield  {author} {\bibinfo {author} {\bibfnamefont {H.}~\bibnamefont
  {Lyu}}\ and\ \bibinfo {author} {\bibfnamefont {Y.}~\bibnamefont {Zhang}},\
  }\bibfield  {title} {\bibinfo {title} {{Spin-orbit-coupling-assisted roton
  softening and superstripes in a Rydberg-dressed Bose-Einstein condensate}},\
  }\href {https://doi.org/10.1103/PhysRevA.102.023327} {\bibfield  {journal}
  {\bibinfo  {journal} {Phys. Rev. A}\ }\textbf {\bibinfo {volume} {102}},\
  \bibinfo {pages} {023327} (\bibinfo {year} {2020})}\BibitemShut {NoStop}%
\bibitem [{\citenamefont {Chen}\ \emph
  {et~al.}(2022{\natexlab{a}})\citenamefont {Chen}, \citenamefont {Lyu},
  \citenamefont {Xu},\ and\ \citenamefont {Zhang}}]{chen2022elementary}%
  \BibitemOpen
  \bibfield  {author} {\bibinfo {author} {\bibfnamefont {Y.}~\bibnamefont
  {Chen}}, \bibinfo {author} {\bibfnamefont {H.}~\bibnamefont {Lyu}}, \bibinfo
  {author} {\bibfnamefont {Y.}~\bibnamefont {Xu}},\ and\ \bibinfo {author}
  {\bibfnamefont {Y.}~\bibnamefont {Zhang}},\ }\bibfield  {title} {\bibinfo
  {title} {{Elementary excitations in a spin--orbit-coupled spin-1
  Bose--Einstein condensate}},\ }\href
  {https://iopscience.iop.org/article/10.1088/1367-2630/ac7fb1/meta} {\bibfield
   {journal} {\bibinfo  {journal} {New J. Phys.}\ }\textbf {\bibinfo {volume}
  {24}},\ \bibinfo {pages} {073041} (\bibinfo {year}
  {2022}{\natexlab{a}})}\BibitemShut {NoStop}%
\bibitem [{\citenamefont {Zhu}\ \emph {et~al.}(2012)\citenamefont {Zhu},
  \citenamefont {Zhang},\ and\ \citenamefont {Wu}}]{zhu2012exotic}%
  \BibitemOpen
  \bibfield  {author} {\bibinfo {author} {\bibfnamefont {Q.}~\bibnamefont
  {Zhu}}, \bibinfo {author} {\bibfnamefont {C.}~\bibnamefont {Zhang}},\ and\
  \bibinfo {author} {\bibfnamefont {B.}~\bibnamefont {Wu}},\ }\bibfield
  {title} {\bibinfo {title} {{Exotic superfluidity in spin-orbit coupled
  Bose-Einstein condensates}},\ }\href
  {https://iopscience.iop.org/article/10.1209/0295-5075/100/50003/meta?casa_token=3X1nJEV0VuYAAAAA:t2Or3z0SNf0QHArSPP6A5PGpivc5VykODIHXSA-fjaaOANv8HJRZdT43a7s3p5ezfRox9iFdQ73wuM6eZgeNJiGysnk}
  {\bibfield  {journal} {\bibinfo  {journal} {Europhys. Lett.}\ }\textbf
  {\bibinfo {volume} {100}},\ \bibinfo {pages} {50003} (\bibinfo {year}
  {2012})}\BibitemShut {NoStop}%
\bibitem [{\citenamefont {Zhang}\ \emph
  {et~al.}(2016{\natexlab{b}})\citenamefont {Zhang}, \citenamefont {Yu},
  \citenamefont {Ng}, \citenamefont {Zhang}, \citenamefont {Pitaevskii},\ and\
  \citenamefont {Stringari}}]{PhysRevA.94.033635}%
  \BibitemOpen
  \bibfield  {author} {\bibinfo {author} {\bibfnamefont {Y.-C.}\ \bibnamefont
  {Zhang}}, \bibinfo {author} {\bibfnamefont {Z.-Q.}\ \bibnamefont {Yu}},
  \bibinfo {author} {\bibfnamefont {T.~K.}\ \bibnamefont {Ng}}, \bibinfo
  {author} {\bibfnamefont {S.}~\bibnamefont {Zhang}}, \bibinfo {author}
  {\bibfnamefont {L.}~\bibnamefont {Pitaevskii}},\ and\ \bibinfo {author}
  {\bibfnamefont {S.}~\bibnamefont {Stringari}},\ }\bibfield  {title} {\bibinfo
  {title} {{Superfluid density of a spin-orbit-coupled Bose gas}},\ }\href
  {https://doi.org/10.1103/PhysRevA.94.033635} {\bibfield  {journal} {\bibinfo
  {journal} {Phys. Rev. A}\ }\textbf {\bibinfo {volume} {94}},\ \bibinfo
  {pages} {033635} (\bibinfo {year} {2016}{\natexlab{b}})}\BibitemShut
  {NoStop}%
\bibitem [{\citenamefont {Yu}(2017)}]{PhysRevA.95.033618}%
  \BibitemOpen
  \bibfield  {author} {\bibinfo {author} {\bibfnamefont {Z.-Q.}\ \bibnamefont
  {Yu}},\ }\bibfield  {title} {\bibinfo {title} {Landau criterion for an
  anisotropic bose-einstein condensate},\ }\href
  {https://doi.org/10.1103/PhysRevA.95.033618} {\bibfield  {journal} {\bibinfo
  {journal} {Phys. Rev. A}\ }\textbf {\bibinfo {volume} {95}},\ \bibinfo
  {pages} {033618} (\bibinfo {year} {2017})}\BibitemShut {NoStop}%
\bibitem [{\citenamefont {Chen}\ \emph
  {et~al.}(2022{\natexlab{b}})\citenamefont {Chen}, \citenamefont {Liu},\ and\
  \citenamefont {Hu}}]{PhysRevA.106.023302}%
  \BibitemOpen
  \bibfield  {author} {\bibinfo {author} {\bibfnamefont {X.-L.}\ \bibnamefont
  {Chen}}, \bibinfo {author} {\bibfnamefont {X.-J.}\ \bibnamefont {Liu}},\ and\
  \bibinfo {author} {\bibfnamefont {H.}~\bibnamefont {Hu}},\ }\bibfield
  {title} {\bibinfo {title} {{Superfluidity of a Raman spin-orbit-coupled Bose
  gas at finite temperature}},\ }\href
  {https://doi.org/10.1103/PhysRevA.106.023302} {\bibfield  {journal} {\bibinfo
   {journal} {Phys. Rev. A}\ }\textbf {\bibinfo {volume} {106}},\ \bibinfo
  {pages} {023302} (\bibinfo {year} {2022}{\natexlab{b}})}\BibitemShut
  {NoStop}%
\bibitem [{\citenamefont {Stringari}(2017)}]{PhysRevLett.118.145302}%
  \BibitemOpen
  \bibfield  {author} {\bibinfo {author} {\bibfnamefont {S.}~\bibnamefont
  {Stringari}},\ }\bibfield  {title} {\bibinfo {title} {{Diffused Vorticity and
  Moment of Inertia of a Spin-Orbit Coupled Bose-Einstein Condensate}},\ }\href
  {https://doi.org/10.1103/PhysRevLett.118.145302} {\bibfield  {journal}
  {\bibinfo  {journal} {Phys. Rev. Lett.}\ }\textbf {\bibinfo {volume} {118}},\
  \bibinfo {pages} {145302} (\bibinfo {year} {2017})}\BibitemShut {NoStop}%
\bibitem [{\citenamefont {Zhang}\ \emph {et~al.}(2012)\citenamefont {Zhang},
  \citenamefont {Ji}, \citenamefont {Chen}, \citenamefont {Zhang},
  \citenamefont {Du}, \citenamefont {Yan}, \citenamefont {Pan}, \citenamefont
  {Zhao}, \citenamefont {Deng}, \citenamefont {Zhai}, \citenamefont {Chen},\
  and\ \citenamefont {Pan}}]{PhysRevLett.109.115301}%
  \BibitemOpen
  \bibfield  {author} {\bibinfo {author} {\bibfnamefont {J.-Y.}\ \bibnamefont
  {Zhang}}, \bibinfo {author} {\bibfnamefont {S.-C.}\ \bibnamefont {Ji}},
  \bibinfo {author} {\bibfnamefont {Z.}~\bibnamefont {Chen}}, \bibinfo {author}
  {\bibfnamefont {L.}~\bibnamefont {Zhang}}, \bibinfo {author} {\bibfnamefont
  {Z.-D.}\ \bibnamefont {Du}}, \bibinfo {author} {\bibfnamefont
  {B.}~\bibnamefont {Yan}}, \bibinfo {author} {\bibfnamefont {G.-S.}\
  \bibnamefont {Pan}}, \bibinfo {author} {\bibfnamefont {B.}~\bibnamefont
  {Zhao}}, \bibinfo {author} {\bibfnamefont {Y.-J.}\ \bibnamefont {Deng}},
  \bibinfo {author} {\bibfnamefont {H.}~\bibnamefont {Zhai}}, \bibinfo {author}
  {\bibfnamefont {S.}~\bibnamefont {Chen}},\ and\ \bibinfo {author}
  {\bibfnamefont {J.-W.}\ \bibnamefont {Pan}},\ }\bibfield  {title} {\bibinfo
  {title} {{Collective Dipole Oscillations of a Spin-Orbit Coupled
  Bose-Einstein Condensate}},\ }\href
  {https://doi.org/10.1103/PhysRevLett.109.115301} {\bibfield  {journal}
  {\bibinfo  {journal} {Phys. Rev. Lett.}\ }\textbf {\bibinfo {volume} {109}},\
  \bibinfo {pages} {115301} (\bibinfo {year} {2012})}\BibitemShut {NoStop}%
\bibitem [{\citenamefont {Chen}\ and\ \citenamefont
  {Zhai}(2012)}]{PhysRevA.86.041604}%
  \BibitemOpen
  \bibfield  {author} {\bibinfo {author} {\bibfnamefont {Z.}~\bibnamefont
  {Chen}}\ and\ \bibinfo {author} {\bibfnamefont {H.}~\bibnamefont {Zhai}},\
  }\bibfield  {title} {\bibinfo {title} {{Collective-mode dynamics in a
  spin-orbit-coupled Bose-Einstein condensate}},\ }\href
  {https://doi.org/10.1103/PhysRevA.86.041604} {\bibfield  {journal} {\bibinfo
  {journal} {Phys. Rev. A}\ }\textbf {\bibinfo {volume} {86}},\ \bibinfo
  {pages} {041604} (\bibinfo {year} {2012})}\BibitemShut {NoStop}%
\bibitem [{\citenamefont {Qu}\ \emph {et~al.}(2013)\citenamefont {Qu},
  \citenamefont {Hamner}, \citenamefont {Gong}, \citenamefont {Zhang},\ and\
  \citenamefont {Engels}}]{PhysRevA.88.021604}%
  \BibitemOpen
  \bibfield  {author} {\bibinfo {author} {\bibfnamefont {C.}~\bibnamefont
  {Qu}}, \bibinfo {author} {\bibfnamefont {C.}~\bibnamefont {Hamner}}, \bibinfo
  {author} {\bibfnamefont {M.}~\bibnamefont {Gong}}, \bibinfo {author}
  {\bibfnamefont {C.}~\bibnamefont {Zhang}},\ and\ \bibinfo {author}
  {\bibfnamefont {P.}~\bibnamefont {Engels}},\ }\bibfield  {title} {\bibinfo
  {title} {{Observation of Zitterbewegung in a spin-orbit-coupled Bose-Einstein
  condensate}},\ }\href {https://doi.org/10.1103/PhysRevA.88.021604} {\bibfield
   {journal} {\bibinfo  {journal} {Phys. Rev. A}\ }\textbf {\bibinfo {volume}
  {88}},\ \bibinfo {pages} {021604} (\bibinfo {year} {2013})}\BibitemShut
  {NoStop}%
\bibitem [{\citenamefont {Olson}\ \emph {et~al.}(2014)\citenamefont {Olson},
  \citenamefont {Wang}, \citenamefont {Niffenegger}, \citenamefont {Li},
  \citenamefont {Greene},\ and\ \citenamefont {Chen}}]{PhysRevA.90.013616}%
  \BibitemOpen
  \bibfield  {author} {\bibinfo {author} {\bibfnamefont {A.~J.}\ \bibnamefont
  {Olson}}, \bibinfo {author} {\bibfnamefont {S.-J.}\ \bibnamefont {Wang}},
  \bibinfo {author} {\bibfnamefont {R.~J.}\ \bibnamefont {Niffenegger}},
  \bibinfo {author} {\bibfnamefont {C.-H.}\ \bibnamefont {Li}}, \bibinfo
  {author} {\bibfnamefont {C.~H.}\ \bibnamefont {Greene}},\ and\ \bibinfo
  {author} {\bibfnamefont {Y.~P.}\ \bibnamefont {Chen}},\ }\bibfield  {title}
  {\bibinfo {title} {{Tunable Landau-Zener transitions in a spin-orbit-coupled
  Bose-Einstein condensate}},\ }\href
  {https://doi.org/10.1103/PhysRevA.90.013616} {\bibfield  {journal} {\bibinfo
  {journal} {Phys. Rev. A}\ }\textbf {\bibinfo {volume} {90}},\ \bibinfo
  {pages} {013616} (\bibinfo {year} {2014})}\BibitemShut {NoStop}%
\bibitem [{\citenamefont {Mardonov}\ \emph
  {et~al.}(2015{\natexlab{a}})\citenamefont {Mardonov}, \citenamefont
  {Modugno},\ and\ \citenamefont {Sherman}}]{PhysRevLett.115.180402}%
  \BibitemOpen
  \bibfield  {author} {\bibinfo {author} {\bibfnamefont {S.}~\bibnamefont
  {Mardonov}}, \bibinfo {author} {\bibfnamefont {M.}~\bibnamefont {Modugno}},\
  and\ \bibinfo {author} {\bibfnamefont {E.~Y.}\ \bibnamefont {Sherman}},\
  }\bibfield  {title} {\bibinfo {title} {{Dynamics of Spin-Orbit Coupled
  Bose-Einstein Condensates in a Random Potential}},\ }\href
  {https://doi.org/10.1103/PhysRevLett.115.180402} {\bibfield  {journal}
  {\bibinfo  {journal} {Phys. Rev. Lett.}\ }\textbf {\bibinfo {volume} {115}},\
  \bibinfo {pages} {180402} (\bibinfo {year} {2015}{\natexlab{a}})}\BibitemShut
  {NoStop}%
\bibitem [{\citenamefont {Mardonov}\ \emph
  {et~al.}(2015{\natexlab{b}})\citenamefont {Mardonov}, \citenamefont
  {Sherman}, \citenamefont {Muga}, \citenamefont {Wang}, \citenamefont {Ban},\
  and\ \citenamefont {Chen}}]{PhysRevA.91.043604}%
  \BibitemOpen
  \bibfield  {author} {\bibinfo {author} {\bibfnamefont {S.}~\bibnamefont
  {Mardonov}}, \bibinfo {author} {\bibfnamefont {E.~Y.}\ \bibnamefont
  {Sherman}}, \bibinfo {author} {\bibfnamefont {J.~G.}\ \bibnamefont {Muga}},
  \bibinfo {author} {\bibfnamefont {H.-W.}\ \bibnamefont {Wang}}, \bibinfo
  {author} {\bibfnamefont {Y.}~\bibnamefont {Ban}},\ and\ \bibinfo {author}
  {\bibfnamefont {X.}~\bibnamefont {Chen}},\ }\bibfield  {title} {\bibinfo
  {title} {{Collapse of spin-orbit-coupled Bose-Einstein condensates}},\ }\href
  {https://doi.org/10.1103/PhysRevA.91.043604} {\bibfield  {journal} {\bibinfo
  {journal} {Phys. Rev. A}\ }\textbf {\bibinfo {volume} {91}},\ \bibinfo
  {pages} {043604} (\bibinfo {year} {2015}{\natexlab{b}})}\BibitemShut
  {NoStop}%
\bibitem [{\citenamefont {Qu}\ \emph {et~al.}(2017)\citenamefont {Qu},
  \citenamefont {Pitaevskii},\ and\ \citenamefont {Stringari}}]{qu2017spin}%
  \BibitemOpen
  \bibfield  {author} {\bibinfo {author} {\bibfnamefont {C.}~\bibnamefont
  {Qu}}, \bibinfo {author} {\bibfnamefont {L.~P.}\ \bibnamefont {Pitaevskii}},\
  and\ \bibinfo {author} {\bibfnamefont {S.}~\bibnamefont {Stringari}},\
  }\bibfield  {title} {\bibinfo {title} {{Spin-orbit-coupling induced
  localization in the expansion of an interacting Bose-Einstein condensate}},\
  }\href {https://iopscience.iop.org/article/10.1088/1367-2630/aa7e8c/meta}
  {\bibfield  {journal} {\bibinfo  {journal} {New J. Phys.}\ }\textbf {\bibinfo
  {volume} {19}},\ \bibinfo {pages} {085006} (\bibinfo {year}
  {2017})}\BibitemShut {NoStop}%
\bibitem [{\citenamefont {Li}\ \emph {et~al.}(2019)\citenamefont {Li},
  \citenamefont {Qu}, \citenamefont {Niffenegger}, \citenamefont {Wang},
  \citenamefont {He}, \citenamefont {Blasing}, \citenamefont {Olson},
  \citenamefont {Greene}, \citenamefont {Lyanda-Geller}, \citenamefont {Zhou}
  \emph {et~al.}}]{li2019spin1}%
  \BibitemOpen
  \bibfield  {author} {\bibinfo {author} {\bibfnamefont {C.-H.}\ \bibnamefont
  {Li}}, \bibinfo {author} {\bibfnamefont {C.}~\bibnamefont {Qu}}, \bibinfo
  {author} {\bibfnamefont {R.~J.}\ \bibnamefont {Niffenegger}}, \bibinfo
  {author} {\bibfnamefont {S.-J.}\ \bibnamefont {Wang}}, \bibinfo {author}
  {\bibfnamefont {M.}~\bibnamefont {He}}, \bibinfo {author} {\bibfnamefont
  {D.~B.}\ \bibnamefont {Blasing}}, \bibinfo {author} {\bibfnamefont {A.~J.}\
  \bibnamefont {Olson}}, \bibinfo {author} {\bibfnamefont {C.~H.}\ \bibnamefont
  {Greene}}, \bibinfo {author} {\bibfnamefont {Y.}~\bibnamefont
  {Lyanda-Geller}}, \bibinfo {author} {\bibfnamefont {Q.}~\bibnamefont {Zhou}},
  \emph {et~al.},\ }\bibfield  {title} {\bibinfo {title} {{Spin current
  generation and relaxation in a quenched spin-orbit-coupled Bose-Einstein
  condensate}},\ }\href {https://www.nature.com/articles/s41467-018-08119-4}
  {\bibfield  {journal} {\bibinfo  {journal} {Nat. Commun.}\ }\textbf {\bibinfo
  {volume} {10}},\ \bibinfo {pages} {375} (\bibinfo {year} {2019})}\BibitemShut
  {NoStop}%
\bibitem [{\citenamefont {Geier}\ \emph {et~al.}(2023)\citenamefont {Geier},
  \citenamefont {Martone}, \citenamefont {Hauke}, \citenamefont {Ketterle},\
  and\ \citenamefont {Stringari}}]{PhysRevLett.130.156001}%
  \BibitemOpen
  \bibfield  {author} {\bibinfo {author} {\bibfnamefont {K.~T.}\ \bibnamefont
  {Geier}}, \bibinfo {author} {\bibfnamefont {G.~I.}\ \bibnamefont {Martone}},
  \bibinfo {author} {\bibfnamefont {P.}~\bibnamefont {Hauke}}, \bibinfo
  {author} {\bibfnamefont {W.}~\bibnamefont {Ketterle}},\ and\ \bibinfo
  {author} {\bibfnamefont {S.}~\bibnamefont {Stringari}},\ }\bibfield  {title}
  {\bibinfo {title} {{Dynamics of Stripe Patterns in Supersolid
  Spin-Orbit-Coupled Bose Gases}},\ }\href
  {https://doi.org/10.1103/PhysRevLett.130.156001} {\bibfield  {journal}
  {\bibinfo  {journal} {Phys. Rev. Lett.}\ }\textbf {\bibinfo {volume} {130}},\
  \bibinfo {pages} {156001} (\bibinfo {year} {2023})}\BibitemShut {NoStop}%
\bibitem [{\citenamefont {Hasan}\ \emph {et~al.}(2022)\citenamefont {Hasan},
  \citenamefont {Polo}, \citenamefont {Pelayo},\ and\ \citenamefont
  {Busch}}]{Hasan2022}%
  \BibitemOpen
  \bibfield  {author} {\bibinfo {author} {\bibfnamefont {M.~S.}\ \bibnamefont
  {Hasan}}, \bibinfo {author} {\bibfnamefont {J.}~\bibnamefont {Polo}},
  \bibinfo {author} {\bibfnamefont {J.~C.}\ \bibnamefont {Pelayo}},\ and\
  \bibinfo {author} {\bibfnamefont {T.}~\bibnamefont {Busch}},\ }\bibfield
  {title} {\bibinfo {title} {{Bloch oscillations in supersolids}},\ }\href
  {https://doi.org/10.1088/1361-6455/ac6ea3} {\bibfield  {journal} {\bibinfo
  {journal} {J. Phys. B: At., Mol. Opt. Phys}\ }\textbf {\bibinfo {volume}
  {55}},\ \bibinfo {pages} {135302} (\bibinfo {year} {2022})}\BibitemShut
  {NoStop}%
\bibitem [{\citenamefont {Qu}\ \emph {et~al.}(2023)\citenamefont {Qu},
  \citenamefont {Li}, \citenamefont {Chen},\ and\ \citenamefont
  {Stringari}}]{PhysRevA.108.053316}%
  \BibitemOpen
  \bibfield  {author} {\bibinfo {author} {\bibfnamefont {C.}~\bibnamefont
  {Qu}}, \bibinfo {author} {\bibfnamefont {C.-H.}\ \bibnamefont {Li}}, \bibinfo
  {author} {\bibfnamefont {Y.~P.}\ \bibnamefont {Chen}},\ and\ \bibinfo
  {author} {\bibfnamefont {S.}~\bibnamefont {Stringari}},\ }\bibfield  {title}
  {\bibinfo {title} {{Scissors modes of a Bose-Einstein condensate in a
  synthetic magnetic field}},\ }\href
  {https://doi.org/10.1103/PhysRevA.108.053316} {\bibfield  {journal} {\bibinfo
   {journal} {Phys. Rev. A}\ }\textbf {\bibinfo {volume} {108}},\ \bibinfo
  {pages} {053316} (\bibinfo {year} {2023})}\BibitemShut {NoStop}%
\bibitem [{\citenamefont {Hamner}\ \emph {et~al.}(2015)\citenamefont {Hamner},
  \citenamefont {Zhang}, \citenamefont {Khamehchi}, \citenamefont {Davis},\
  and\ \citenamefont {Engels}}]{PhysRevLett.114.070401}%
  \BibitemOpen
  \bibfield  {author} {\bibinfo {author} {\bibfnamefont {C.}~\bibnamefont
  {Hamner}}, \bibinfo {author} {\bibfnamefont {Y.}~\bibnamefont {Zhang}},
  \bibinfo {author} {\bibfnamefont {M.~A.}\ \bibnamefont {Khamehchi}}, \bibinfo
  {author} {\bibfnamefont {M.~J.}\ \bibnamefont {Davis}},\ and\ \bibinfo
  {author} {\bibfnamefont {P.}~\bibnamefont {Engels}},\ }\bibfield  {title}
  {\bibinfo {title} {{Spin-Orbit-Coupled Bose-Einstein Condensates in a
  One-Dimensional Optical Lattice}},\ }\href
  {https://doi.org/10.1103/PhysRevLett.114.070401} {\bibfield  {journal}
  {\bibinfo  {journal} {Phys. Rev. Lett.}\ }\textbf {\bibinfo {volume} {114}},\
  \bibinfo {pages} {070401} (\bibinfo {year} {2015})}\BibitemShut {NoStop}%
\bibitem [{\citenamefont {Cai}\ \emph {et~al.}(2012)\citenamefont {Cai},
  \citenamefont {Zhou},\ and\ \citenamefont {Wu}}]{PhysRevA.85.061605}%
  \BibitemOpen
  \bibfield  {author} {\bibinfo {author} {\bibfnamefont {Z.}~\bibnamefont
  {Cai}}, \bibinfo {author} {\bibfnamefont {X.}~\bibnamefont {Zhou}},\ and\
  \bibinfo {author} {\bibfnamefont {C.}~\bibnamefont {Wu}},\ }\bibfield
  {title} {\bibinfo {title} {Magnetic phases of bosons with synthetic
  spin-orbit coupling in optical lattices},\ }\href
  {https://doi.org/10.1103/PhysRevA.85.061605} {\bibfield  {journal} {\bibinfo
  {journal} {Phys. Rev. A}\ }\textbf {\bibinfo {volume} {85}},\ \bibinfo
  {pages} {061605} (\bibinfo {year} {2012})}\BibitemShut {NoStop}%
\bibitem [{\citenamefont {Radi\ifmmode~\acute{c}\else \'{c}\fi{}}\ \emph
  {et~al.}(2012)\citenamefont {Radi\ifmmode~\acute{c}\else \'{c}\fi{}},
  \citenamefont {Di~Ciolo}, \citenamefont {Sun},\ and\ \citenamefont
  {Galitski}}]{PhysRevLett.109.085303}%
  \BibitemOpen
  \bibfield  {author} {\bibinfo {author} {\bibfnamefont {J.}~\bibnamefont
  {Radi\ifmmode~\acute{c}\else \'{c}\fi{}}}, \bibinfo {author} {\bibfnamefont
  {A.}~\bibnamefont {Di~Ciolo}}, \bibinfo {author} {\bibfnamefont
  {K.}~\bibnamefont {Sun}},\ and\ \bibinfo {author} {\bibfnamefont
  {V.}~\bibnamefont {Galitski}},\ }\bibfield  {title} {\bibinfo {title} {Exotic
  quantum spin models in spin-orbit-coupled mott insulators},\ }\href
  {https://doi.org/10.1103/PhysRevLett.109.085303} {\bibfield  {journal}
  {\bibinfo  {journal} {Phys. Rev. Lett.}\ }\textbf {\bibinfo {volume} {109}},\
  \bibinfo {pages} {085303} (\bibinfo {year} {2012})}\BibitemShut {NoStop}%
\bibitem [{\citenamefont {Cole}\ \emph {et~al.}(2012)\citenamefont {Cole},
  \citenamefont {Zhang}, \citenamefont {Paramekanti},\ and\ \citenamefont
  {Trivedi}}]{PhysRevLett.109.085302}%
  \BibitemOpen
  \bibfield  {author} {\bibinfo {author} {\bibfnamefont {W.~S.}\ \bibnamefont
  {Cole}}, \bibinfo {author} {\bibfnamefont {S.}~\bibnamefont {Zhang}},
  \bibinfo {author} {\bibfnamefont {A.}~\bibnamefont {Paramekanti}},\ and\
  \bibinfo {author} {\bibfnamefont {N.}~\bibnamefont {Trivedi}},\ }\bibfield
  {title} {\bibinfo {title} {Bose-hubbard models with synthetic spin-orbit
  coupling: Mott insulators, spin textures, and superfluidity},\ }\href
  {https://doi.org/10.1103/PhysRevLett.109.085302} {\bibfield  {journal}
  {\bibinfo  {journal} {Phys. Rev. Lett.}\ }\textbf {\bibinfo {volume} {109}},\
  \bibinfo {pages} {085302} (\bibinfo {year} {2012})}\BibitemShut {NoStop}%
\bibitem [{\citenamefont {Zhao}\ \emph {et~al.}(2014)\citenamefont {Zhao},
  \citenamefont {Hu}, \citenamefont {Chang}, \citenamefont {Zheng},
  \citenamefont {Zhang},\ and\ \citenamefont {Wang}}]{PhysRevB.90.085117}%
  \BibitemOpen
  \bibfield  {author} {\bibinfo {author} {\bibfnamefont {J.}~\bibnamefont
  {Zhao}}, \bibinfo {author} {\bibfnamefont {S.}~\bibnamefont {Hu}}, \bibinfo
  {author} {\bibfnamefont {J.}~\bibnamefont {Chang}}, \bibinfo {author}
  {\bibfnamefont {F.}~\bibnamefont {Zheng}}, \bibinfo {author} {\bibfnamefont
  {P.}~\bibnamefont {Zhang}},\ and\ \bibinfo {author} {\bibfnamefont
  {X.}~\bibnamefont {Wang}},\ }\bibfield  {title} {\bibinfo {title} {Evolution
  of magnetic structure driven by synthetic spin-orbit coupling in a
  two-component bose-hubbard model},\ }\href
  {https://doi.org/10.1103/PhysRevB.90.085117} {\bibfield  {journal} {\bibinfo
  {journal} {Phys. Rev. B}\ }\textbf {\bibinfo {volume} {90}},\ \bibinfo
  {pages} {085117} (\bibinfo {year} {2014})}\BibitemShut {NoStop}%
\bibitem [{\citenamefont {Gong}\ \emph {et~al.}(2015)\citenamefont {Gong},
  \citenamefont {Qian}, \citenamefont {Yan}, \citenamefont {Scarola},\ and\
  \citenamefont {Zhang}}]{gong2015dz}%
  \BibitemOpen
  \bibfield  {author} {\bibinfo {author} {\bibfnamefont {M.}~\bibnamefont
  {Gong}}, \bibinfo {author} {\bibfnamefont {Y.}~\bibnamefont {Qian}}, \bibinfo
  {author} {\bibfnamefont {M.}~\bibnamefont {Yan}}, \bibinfo {author}
  {\bibfnamefont {V.~W.}\ \bibnamefont {Scarola}},\ and\ \bibinfo {author}
  {\bibfnamefont {C.}~\bibnamefont {Zhang}},\ }\bibfield  {title} {\bibinfo
  {title} {Dzyaloshinskii-moriya interaction and spiral order in spin-orbit
  coupled optical lattices},\ }\href {https://doi.org/10.1038/srep10050}
  {\bibfield  {journal} {\bibinfo  {journal} {Scientific Reports}\ }\textbf
  {\bibinfo {volume} {5}},\ \bibinfo {pages} {10050} (\bibinfo {year}
  {2015})}\BibitemShut {NoStop}%
\bibitem [{\citenamefont {Zhang}\ \emph {et~al.}(2019)\citenamefont {Zhang},
  \citenamefont {Ke},\ and\ \citenamefont {Lee}}]{PhysRevB.100.224420}%
  \BibitemOpen
  \bibfield  {author} {\bibinfo {author} {\bibfnamefont {L.}~\bibnamefont
  {Zhang}}, \bibinfo {author} {\bibfnamefont {Y.}~\bibnamefont {Ke}},\ and\
  \bibinfo {author} {\bibfnamefont {C.}~\bibnamefont {Lee}},\ }\bibfield
  {title} {\bibinfo {title} {{Magnetic phase transitions of insulating
  spin-orbit coupled Bose atoms in one-dimensional optical lattices}},\ }\href
  {https://doi.org/10.1103/PhysRevB.100.224420} {\bibfield  {journal} {\bibinfo
   {journal} {Phys. Rev. B}\ }\textbf {\bibinfo {volume} {100}},\ \bibinfo
  {pages} {224420} (\bibinfo {year} {2019})}\BibitemShut {NoStop}%
\bibitem [{\citenamefont {Yamamoto}\ \emph {et~al.}(2017)\citenamefont
  {Yamamoto}, \citenamefont {Spielman},\ and\ \citenamefont {S\'a~de
  Melo}}]{PhysRevA.96.061603}%
  \BibitemOpen
  \bibfield  {author} {\bibinfo {author} {\bibfnamefont {D.}~\bibnamefont
  {Yamamoto}}, \bibinfo {author} {\bibfnamefont {I.~B.}\ \bibnamefont
  {Spielman}},\ and\ \bibinfo {author} {\bibfnamefont {C.~A.~R.}\ \bibnamefont
  {S\'a~de Melo}},\ }\bibfield  {title} {\bibinfo {title} {{Quantum phases of
  two-component bosons with spin-orbit coupling in optical lattices}},\ }\href
  {https://doi.org/10.1103/PhysRevA.96.061603} {\bibfield  {journal} {\bibinfo
  {journal} {Phys. Rev. A}\ }\textbf {\bibinfo {volume} {96}},\ \bibinfo
  {pages} {061603} (\bibinfo {year} {2017})}\BibitemShut {NoStop}%
\bibitem [{\citenamefont {Zhang}\ and\ \citenamefont
  {Zhang}(2013)}]{PhysRevA.87.023611}%
  \BibitemOpen
  \bibfield  {author} {\bibinfo {author} {\bibfnamefont {Y.}~\bibnamefont
  {Zhang}}\ and\ \bibinfo {author} {\bibfnamefont {C.}~\bibnamefont {Zhang}},\
  }\bibfield  {title} {\bibinfo {title} {{Bose-Einstein condensates in
  spin-orbit-coupled optical lattices: Flat bands and superfluidity}},\ }\href
  {https://doi.org/10.1103/PhysRevA.87.023611} {\bibfield  {journal} {\bibinfo
  {journal} {Phys. Rev. A}\ }\textbf {\bibinfo {volume} {87}},\ \bibinfo
  {pages} {023611} (\bibinfo {year} {2013})}\BibitemShut {NoStop}%
\bibitem [{\citenamefont {Li}\ \emph {et~al.}(2015)\citenamefont {Li},
  \citenamefont {Chen}, \citenamefont {Chen}, \citenamefont {Hu}, \citenamefont
  {Zhang},\ and\ \citenamefont {Liang}}]{PhysRevA.91.023629}%
  \BibitemOpen
  \bibfield  {author} {\bibinfo {author} {\bibfnamefont {W.}~\bibnamefont
  {Li}}, \bibinfo {author} {\bibfnamefont {L.}~\bibnamefont {Chen}}, \bibinfo
  {author} {\bibfnamefont {Z.}~\bibnamefont {Chen}}, \bibinfo {author}
  {\bibfnamefont {Y.}~\bibnamefont {Hu}}, \bibinfo {author} {\bibfnamefont
  {Z.}~\bibnamefont {Zhang}},\ and\ \bibinfo {author} {\bibfnamefont
  {Z.}~\bibnamefont {Liang}},\ }\bibfield  {title} {\bibinfo {title} {{Probing
  the flat band of optically trapped spin-orbital-coupled Bose gases using
  Bragg spectroscopy}},\ }\href {https://doi.org/10.1103/PhysRevA.91.023629}
  {\bibfield  {journal} {\bibinfo  {journal} {Phys. Rev. A}\ }\textbf {\bibinfo
  {volume} {91}},\ \bibinfo {pages} {023629} (\bibinfo {year}
  {2015})}\BibitemShut {NoStop}%
\bibitem [{\citenamefont {Salerno}\ \emph {et~al.}(2016)\citenamefont
  {Salerno}, \citenamefont {Abdullaev}, \citenamefont {Gammal},\ and\
  \citenamefont {Tomio}}]{PhysRevA.94.043602}%
  \BibitemOpen
  \bibfield  {author} {\bibinfo {author} {\bibfnamefont {M.}~\bibnamefont
  {Salerno}}, \bibinfo {author} {\bibfnamefont {F.~K.}\ \bibnamefont
  {Abdullaev}}, \bibinfo {author} {\bibfnamefont {A.}~\bibnamefont {Gammal}},\
  and\ \bibinfo {author} {\bibfnamefont {L.}~\bibnamefont {Tomio}},\ }\bibfield
   {title} {\bibinfo {title} {{Tunable spin-orbit-coupled Bose-Einstein
  condensates in deep optical lattices}},\ }\href
  {https://doi.org/10.1103/PhysRevA.94.043602} {\bibfield  {journal} {\bibinfo
  {journal} {Phys. Rev. A}\ }\textbf {\bibinfo {volume} {94}},\ \bibinfo
  {pages} {043602} (\bibinfo {year} {2016})}\BibitemShut {NoStop}%
\bibitem [{\citenamefont {Lin}\ \emph {et~al.}(2014)\citenamefont {Lin},
  \citenamefont {Zhang},\ and\ \citenamefont
  {Scarola}}]{PhysRevLett.112.110404}%
  \BibitemOpen
  \bibfield  {author} {\bibinfo {author} {\bibfnamefont {F.}~\bibnamefont
  {Lin}}, \bibinfo {author} {\bibfnamefont {C.}~\bibnamefont {Zhang}},\ and\
  \bibinfo {author} {\bibfnamefont {V.~W.}\ \bibnamefont {Scarola}},\
  }\bibfield  {title} {\bibinfo {title} {Emergent kinetics and fractionalized
  charge in 1d spin-orbit coupled flatband optical lattices},\ }\href
  {https://doi.org/10.1103/PhysRevLett.112.110404} {\bibfield  {journal}
  {\bibinfo  {journal} {Phys. Rev. Lett.}\ }\textbf {\bibinfo {volume} {112}},\
  \bibinfo {pages} {110404} (\bibinfo {year} {2014})}\BibitemShut {NoStop}%
\bibitem [{\citenamefont {Hui}\ \emph {et~al.}(2017)\citenamefont {Hui},
  \citenamefont {Zhang}, \citenamefont {Zhang},\ and\ \citenamefont
  {Scarola}}]{PhysRevA.95.033603}%
  \BibitemOpen
  \bibfield  {author} {\bibinfo {author} {\bibfnamefont {H.-Y.}\ \bibnamefont
  {Hui}}, \bibinfo {author} {\bibfnamefont {Y.}~\bibnamefont {Zhang}}, \bibinfo
  {author} {\bibfnamefont {C.}~\bibnamefont {Zhang}},\ and\ \bibinfo {author}
  {\bibfnamefont {V.~W.}\ \bibnamefont {Scarola}},\ }\bibfield  {title}
  {\bibinfo {title} {Superfluidity in the absence of kinetics in
  spin-orbit-coupled optical lattices},\ }\href
  {https://doi.org/10.1103/PhysRevA.95.033603} {\bibfield  {journal} {\bibinfo
  {journal} {Phys. Rev. A}\ }\textbf {\bibinfo {volume} {95}},\ \bibinfo
  {pages} {033603} (\bibinfo {year} {2017})}\BibitemShut {NoStop}%
\bibitem [{\citenamefont {Chen}\ and\ \citenamefont
  {Liang}(2016)}]{PhysRevA.93.013601}%
  \BibitemOpen
  \bibfield  {author} {\bibinfo {author} {\bibfnamefont {Z.}~\bibnamefont
  {Chen}}\ and\ \bibinfo {author} {\bibfnamefont {Z.}~\bibnamefont {Liang}},\
  }\bibfield  {title} {\bibinfo {title} {{Ground-state phase diagram of a
  spin-orbit-coupled bosonic superfluid in an optical lattice}},\ }\href
  {https://doi.org/10.1103/PhysRevA.93.013601} {\bibfield  {journal} {\bibinfo
  {journal} {Phys. Rev. A}\ }\textbf {\bibinfo {volume} {93}},\ \bibinfo
  {pages} {013601} (\bibinfo {year} {2016})}\BibitemShut {NoStop}%
\bibitem [{\citenamefont {Martone}\ \emph {et~al.}(2016)\citenamefont
  {Martone}, \citenamefont {Ozawa}, \citenamefont {Qu},\ and\ \citenamefont
  {Stringari}}]{PhysRevA.94.043629}%
  \BibitemOpen
  \bibfield  {author} {\bibinfo {author} {\bibfnamefont {G.~I.}\ \bibnamefont
  {Martone}}, \bibinfo {author} {\bibfnamefont {T.}~\bibnamefont {Ozawa}},
  \bibinfo {author} {\bibfnamefont {C.}~\bibnamefont {Qu}},\ and\ \bibinfo
  {author} {\bibfnamefont {S.}~\bibnamefont {Stringari}},\ }\bibfield  {title}
  {\bibinfo {title} {{Optical-lattice-assisted magnetic phase transition in a
  spin-orbit-coupled Bose-Einstein condensate}},\ }\href
  {https://doi.org/10.1103/PhysRevA.94.043629} {\bibfield  {journal} {\bibinfo
  {journal} {Phys. Rev. A}\ }\textbf {\bibinfo {volume} {94}},\ \bibinfo
  {pages} {043629} (\bibinfo {year} {2016})}\BibitemShut {NoStop}%
\bibitem [{\citenamefont {Martone}(2017)}]{martone2017quantum1}%
  \BibitemOpen
  \bibfield  {author} {\bibinfo {author} {\bibfnamefont {G.~I.}\ \bibnamefont
  {Martone}},\ }\bibfield  {title} {\bibinfo {title} {{Quantum Phases and
  Collective Excitations of a Spin-Orbit-Coupled Bose--Einstein Condensate in a
  One-Dimensional Optical Lattice}},\ }\href
  {https://link.springer.com/article/10.1007/s10909-017-1816-9} {\bibfield
  {journal} {\bibinfo  {journal} {J. Low Temp. Phys.}\ }\textbf {\bibinfo
  {volume} {189}},\ \bibinfo {pages} {262} (\bibinfo {year}
  {2017})}\BibitemShut {NoStop}%
\bibitem [{\citenamefont {Pan}\ \emph {et~al.}(2016)\citenamefont {Pan},
  \citenamefont {Zhang}, \citenamefont {Yi},\ and\ \citenamefont
  {Guo}}]{PhysRevA.94.043619}%
  \BibitemOpen
  \bibfield  {author} {\bibinfo {author} {\bibfnamefont {J.-S.}\ \bibnamefont
  {Pan}}, \bibinfo {author} {\bibfnamefont {W.}~\bibnamefont {Zhang}}, \bibinfo
  {author} {\bibfnamefont {W.}~\bibnamefont {Yi}},\ and\ \bibinfo {author}
  {\bibfnamefont {G.-C.}\ \bibnamefont {Guo}},\ }\bibfield  {title} {\bibinfo
  {title} {{Bose-Einstein condensate in an optical lattice with Raman-assisted
  two-dimensional spin-orbit coupling}},\ }\href
  {https://doi.org/10.1103/PhysRevA.94.043619} {\bibfield  {journal} {\bibinfo
  {journal} {Phys. Rev. A}\ }\textbf {\bibinfo {volume} {94}},\ \bibinfo
  {pages} {043619} (\bibinfo {year} {2016})}\BibitemShut {NoStop}%
\bibitem [{\citenamefont {Li}\ \emph {et~al.}(2020)\citenamefont {Li},
  \citenamefont {Wang}, \citenamefont {Li}, \citenamefont {Cui},\ and\
  \citenamefont {Liu}}]{PhysRevA.102.033328}%
  \BibitemOpen
  \bibfield  {author} {\bibinfo {author} {\bibfnamefont {S.}~\bibnamefont
  {Li}}, \bibinfo {author} {\bibfnamefont {H.}~\bibnamefont {Wang}}, \bibinfo
  {author} {\bibfnamefont {F.}~\bibnamefont {Li}}, \bibinfo {author}
  {\bibfnamefont {X.}~\bibnamefont {Cui}},\ and\ \bibinfo {author}
  {\bibfnamefont {B.}~\bibnamefont {Liu}},\ }\bibfield  {title} {\bibinfo
  {title} {{Commensurate-incommensurate supersolid ground state of a
  spin-orbit-coupled Bose-Einstein condensate in one-dimensional optical
  lattices}},\ }\href {https://doi.org/10.1103/PhysRevA.102.033328} {\bibfield
  {journal} {\bibinfo  {journal} {Phys. Rev. A}\ }\textbf {\bibinfo {volume}
  {102}},\ \bibinfo {pages} {033328} (\bibinfo {year} {2020})}\BibitemShut
  {NoStop}%
\bibitem [{\citenamefont {M\ae{}land}\ \emph {et~al.}(2020)\citenamefont
  {M\ae{}land}, \citenamefont {Janss\o{}nn}, \citenamefont {Rygh},\ and\
  \citenamefont {Sudb\o{}}}]{PhysRevA.102.053318}%
  \BibitemOpen
  \bibfield  {author} {\bibinfo {author} {\bibfnamefont {K.}~\bibnamefont
  {M\ae{}land}}, \bibinfo {author} {\bibfnamefont {A.~T.~G.}\ \bibnamefont
  {Janss\o{}nn}}, \bibinfo {author} {\bibfnamefont {J.~H.}\ \bibnamefont
  {Rygh}},\ and\ \bibinfo {author} {\bibfnamefont {A.}~\bibnamefont
  {Sudb\o{}}},\ }\bibfield  {title} {\bibinfo {title} {{Plane- and stripe-wave
  phases of a spin-orbit-coupled Bose-Einstein condensate in an optical lattice
  with a Zeeman field}},\ }\href {https://doi.org/10.1103/PhysRevA.102.053318}
  {\bibfield  {journal} {\bibinfo  {journal} {Phys. Rev. A}\ }\textbf {\bibinfo
  {volume} {102}},\ \bibinfo {pages} {053318} (\bibinfo {year}
  {2020})}\BibitemShut {NoStop}%
\bibitem [{\citenamefont {Li}\ \emph {et~al.}(2021)\citenamefont {Li},
  \citenamefont {Luo}, \citenamefont {Hou},\ and\ \citenamefont
  {Zhang}}]{PhysRevA.104.023311}%
  \BibitemOpen
  \bibfield  {author} {\bibinfo {author} {\bibfnamefont {G.-Q.}\ \bibnamefont
  {Li}}, \bibinfo {author} {\bibfnamefont {X.-W.}\ \bibnamefont {Luo}},
  \bibinfo {author} {\bibfnamefont {J.}~\bibnamefont {Hou}},\ and\ \bibinfo
  {author} {\bibfnamefont {C.}~\bibnamefont {Zhang}},\ }\bibfield  {title}
  {\bibinfo {title} {{Pseudo-Goldstone excitations in a striped Bose-Einstein
  condensate}},\ }\href {https://doi.org/10.1103/PhysRevA.104.023311}
  {\bibfield  {journal} {\bibinfo  {journal} {Phys. Rev. A}\ }\textbf {\bibinfo
  {volume} {104}},\ \bibinfo {pages} {023311} (\bibinfo {year}
  {2021})}\BibitemShut {NoStop}%
\bibitem [{\citenamefont {Guo}\ and\ \citenamefont
  {Zhu}(2024)}]{10.1088/1367-2630/ad98b5}%
  \BibitemOpen
  \bibfield  {author} {\bibinfo {author} {\bibfnamefont {Z.-H.}\ \bibnamefont
  {Guo}}\ and\ \bibinfo {author} {\bibfnamefont {Q.}~\bibnamefont {Zhu}},\
  }\bibfield  {title} {\bibinfo {title} {$\mathbb{Z}_n$ symmetry broken
  supersolid in spin-orbit-coupled bose-einstein condensates},\ }\href
  {http://iopscience.iop.org/article/10.1088/1367-2630/ad98b5} {\bibfield
  {journal} {\bibinfo  {journal} {New Journal of Physics}\ } (\bibinfo {year}
  {2024})}\BibitemShut {NoStop}%
\bibitem [{\citenamefont {Kartashov}\ \emph {et~al.}(2016)\citenamefont
  {Kartashov}, \citenamefont {Konotop}, \citenamefont {Zezyulin},\ and\
  \citenamefont {Torner}}]{PhysRevLett.117.215301}%
  \BibitemOpen
  \bibfield  {author} {\bibinfo {author} {\bibfnamefont {Y.~V.}\ \bibnamefont
  {Kartashov}}, \bibinfo {author} {\bibfnamefont {V.~V.}\ \bibnamefont
  {Konotop}}, \bibinfo {author} {\bibfnamefont {D.~A.}\ \bibnamefont
  {Zezyulin}},\ and\ \bibinfo {author} {\bibfnamefont {L.}~\bibnamefont
  {Torner}},\ }\bibfield  {title} {\bibinfo {title} {{Bloch Oscillations in
  Optical and Zeeman Lattices in the Presence of Spin-Orbit Coupling}},\ }\href
  {https://doi.org/10.1103/PhysRevLett.117.215301} {\bibfield  {journal}
  {\bibinfo  {journal} {Phys. Rev. Lett.}\ }\textbf {\bibinfo {volume} {117}},\
  \bibinfo {pages} {215301} (\bibinfo {year} {2016})}\BibitemShut {NoStop}%
\bibitem [{\citenamefont {Ji}\ \emph {et~al.}(2019)\citenamefont {Ji},
  \citenamefont {Zhang}, \citenamefont {Zhang},\ and\ \citenamefont
  {Zhou}}]{PhysRevA.99.023604}%
  \BibitemOpen
  \bibfield  {author} {\bibinfo {author} {\bibfnamefont {W.}~\bibnamefont
  {Ji}}, \bibinfo {author} {\bibfnamefont {K.}~\bibnamefont {Zhang}}, \bibinfo
  {author} {\bibfnamefont {W.}~\bibnamefont {Zhang}},\ and\ \bibinfo {author}
  {\bibfnamefont {L.}~\bibnamefont {Zhou}},\ }\bibfield  {title} {\bibinfo
  {title} {{Bloch oscillations of spin-orbit-coupled cold atoms in an optical
  lattice and spin-current generation}},\ }\href
  {https://doi.org/10.1103/PhysRevA.99.023604} {\bibfield  {journal} {\bibinfo
  {journal} {Phys. Rev. A}\ }\textbf {\bibinfo {volume} {99}},\ \bibinfo
  {pages} {023604} (\bibinfo {year} {2019})}\BibitemShut {NoStop}%
\bibitem [{\citenamefont {Zhang}\ \emph
  {et~al.}(2022{\natexlab{a}})\citenamefont {Zhang}, \citenamefont {Jiao},
  \citenamefont {Zhang},\ and\ \citenamefont {Xue}}]{zhang2022bloch}%
  \BibitemOpen
  \bibfield  {author} {\bibinfo {author} {\bibfnamefont {Y.-C.}\ \bibnamefont
  {Zhang}}, \bibinfo {author} {\bibfnamefont {C.}~\bibnamefont {Jiao}},
  \bibinfo {author} {\bibfnamefont {A.-X.}\ \bibnamefont {Zhang}},\ and\
  \bibinfo {author} {\bibfnamefont {J.-K.}\ \bibnamefont {Xue}},\ }\bibfield
  {title} {\bibinfo {title} {{The Bloch oscillations of spin-orbit coupled
  Bose-Einstein condensates in deep optical lattices}},\ }\href
  {https://iopscience.iop.org/article/10.1209/0295-5075/ac724f/meta?casa_token=yQbbugh1k9sAAAAA:ldxSvMd5u4-PDeWM-OdY1PJllyJEmuiwIoDaQy3goVpu7RXkBsDWNwjUUPYX5jABk3yW6cgWY3H655mng0YhWomLiqw}
  {\bibfield  {journal} {\bibinfo  {journal} {Europhys. Lett.}\ }\textbf
  {\bibinfo {volume} {138}},\ \bibinfo {pages} {55006} (\bibinfo {year}
  {2022}{\natexlab{a}})}\BibitemShut {NoStop}%
\bibitem [{\citenamefont {Toniolo}\ and\ \citenamefont
  {Linder}(2014)}]{PhysRevA.89.061605}%
  \BibitemOpen
  \bibfield  {author} {\bibinfo {author} {\bibfnamefont {D.}~\bibnamefont
  {Toniolo}}\ and\ \bibinfo {author} {\bibfnamefont {J.}~\bibnamefont
  {Linder}},\ }\bibfield  {title} {\bibinfo {title} {{Superfluidity breakdown
  and multiple roton gaps in spin-orbit-coupled Bose-Einstein condensates in an
  optical lattice}},\ }\href {https://doi.org/10.1103/PhysRevA.89.061605}
  {\bibfield  {journal} {\bibinfo  {journal} {Phys. Rev. A}\ }\textbf {\bibinfo
  {volume} {89}},\ \bibinfo {pages} {061605} (\bibinfo {year}
  {2014})}\BibitemShut {NoStop}%
\bibitem [{\citenamefont {Luo}\ \emph {et~al.}(2021)\citenamefont {Luo},
  \citenamefont {Hu}, \citenamefont {Zeng}, \citenamefont {Luo}, \citenamefont
  {Yang}, \citenamefont {Xiao}, \citenamefont {Li},\ and\ \citenamefont
  {Chen}}]{PhysRevA.103.063324}%
  \BibitemOpen
  \bibfield  {author} {\bibinfo {author} {\bibfnamefont {X.}~\bibnamefont
  {Luo}}, \bibinfo {author} {\bibfnamefont {Z.}~\bibnamefont {Hu}}, \bibinfo
  {author} {\bibfnamefont {Z.-Y.}\ \bibnamefont {Zeng}}, \bibinfo {author}
  {\bibfnamefont {Y.}~\bibnamefont {Luo}}, \bibinfo {author} {\bibfnamefont
  {B.}~\bibnamefont {Yang}}, \bibinfo {author} {\bibfnamefont {J.}~\bibnamefont
  {Xiao}}, \bibinfo {author} {\bibfnamefont {L.}~\bibnamefont {Li}},\ and\
  \bibinfo {author} {\bibfnamefont {A.-X.}\ \bibnamefont {Chen}},\ }\bibfield
  {title} {\bibinfo {title} {{Analytical results for the superflow of
  spin-orbit-coupled Bose-Einstein condensates in optical lattices}},\ }\href
  {https://doi.org/10.1103/PhysRevA.103.063324} {\bibfield  {journal} {\bibinfo
   {journal} {Phys. Rev. A}\ }\textbf {\bibinfo {volume} {103}},\ \bibinfo
  {pages} {063324} (\bibinfo {year} {2021})}\BibitemShut {NoStop}%
\bibitem [{\citenamefont {Orso}(2017)}]{PhysRevLett.118.105301}%
  \BibitemOpen
  \bibfield  {author} {\bibinfo {author} {\bibfnamefont {G.}~\bibnamefont
  {Orso}},\ }\bibfield  {title} {\bibinfo {title} {Anderson transition of cold
  atoms with synthetic spin-orbit coupling in two-dimensional speckle
  potentials},\ }\href {https://doi.org/10.1103/PhysRevLett.118.105301}
  {\bibfield  {journal} {\bibinfo  {journal} {Phys. Rev. Lett.}\ }\textbf
  {\bibinfo {volume} {118}},\ \bibinfo {pages} {105301} (\bibinfo {year}
  {2017})}\BibitemShut {NoStop}%
\bibitem [{\citenamefont {Zhang}\ \emph
  {et~al.}(2022{\natexlab{b}})\citenamefont {Zhang}, \citenamefont {Liu},\ and\
  \citenamefont {Zhang}}]{HuanZhang70305}%
  \BibitemOpen
  \bibfield  {author} {\bibinfo {author} {\bibfnamefont {H.}~\bibnamefont
  {Zhang}}, \bibinfo {author} {\bibfnamefont {S.}~\bibnamefont {Liu}},\ and\
  \bibinfo {author} {\bibfnamefont {Y.}~\bibnamefont {Zhang}},\ }\bibfield
  {title} {\bibinfo {title} {Anderson localization of a spin-orbit coupled
  bose-einstein condensate in disorder potential},\ }\href
  {https://doi.org/10.1088/1674-1056/ac538d} {\bibfield  {journal} {\bibinfo
  {journal} {Chinese Physics B}\ }\textbf {\bibinfo {volume} {31}},\ \bibinfo
  {eid} {070305} (\bibinfo {year} {2022}{\natexlab{b}})}\BibitemShut {NoStop}%
\bibitem [{\citenamefont {Alluf}\ and\ \citenamefont {S\'a~de
  Melo}(2023)}]{PhysRevA.107.033312}%
  \BibitemOpen
  \bibfield  {author} {\bibinfo {author} {\bibfnamefont {B.}~\bibnamefont
  {Alluf}}\ and\ \bibinfo {author} {\bibfnamefont {C.~A.~R.}\ \bibnamefont
  {S\'a~de Melo}},\ }\bibfield  {title} {\bibinfo {title} {Controlling anderson
  localization of a bose-einstein condensate via spin-orbit coupling and rabi
  fields in bichromatic lattices},\ }\href
  {https://doi.org/10.1103/PhysRevA.107.033312} {\bibfield  {journal} {\bibinfo
   {journal} {Phys. Rev. A}\ }\textbf {\bibinfo {volume} {107}},\ \bibinfo
  {pages} {033312} (\bibinfo {year} {2023})}\BibitemShut {NoStop}%
\bibitem [{\citenamefont {Bersano}\ \emph {et~al.}(2019)\citenamefont
  {Bersano}, \citenamefont {Hou}, \citenamefont {Mossman}, \citenamefont
  {Gokhroo}, \citenamefont {Luo}, \citenamefont {Sun}, \citenamefont {Zhang},\
  and\ \citenamefont {Engels}}]{PhysRevA.99.051602}%
  \BibitemOpen
  \bibfield  {author} {\bibinfo {author} {\bibfnamefont {T.~M.}\ \bibnamefont
  {Bersano}}, \bibinfo {author} {\bibfnamefont {J.}~\bibnamefont {Hou}},
  \bibinfo {author} {\bibfnamefont {S.}~\bibnamefont {Mossman}}, \bibinfo
  {author} {\bibfnamefont {V.}~\bibnamefont {Gokhroo}}, \bibinfo {author}
  {\bibfnamefont {X.-W.}\ \bibnamefont {Luo}}, \bibinfo {author} {\bibfnamefont
  {K.}~\bibnamefont {Sun}}, \bibinfo {author} {\bibfnamefont {C.}~\bibnamefont
  {Zhang}},\ and\ \bibinfo {author} {\bibfnamefont {P.}~\bibnamefont
  {Engels}},\ }\bibfield  {title} {\bibinfo {title} {{Experimental realization
  of a long-lived striped Bose-Einstein condensate induced by momentum-space
  hopping}},\ }\href {https://doi.org/10.1103/PhysRevA.99.051602} {\bibfield
  {journal} {\bibinfo  {journal} {Phys. Rev. A}\ }\textbf {\bibinfo {volume}
  {99}},\ \bibinfo {pages} {051602} (\bibinfo {year} {2019})}\BibitemShut
  {NoStop}%
\bibitem [{\citenamefont {Mukhopadhyay}\ \emph {et~al.}(2024)\citenamefont
  {Mukhopadhyay}, \citenamefont {Luo}, \citenamefont {Schimelfenig},
  \citenamefont {Ome}, \citenamefont {Mossman}, \citenamefont {Zhang},\ and\
  \citenamefont {Engels}}]{PhysRevLett.132.233403}%
  \BibitemOpen
  \bibfield  {author} {\bibinfo {author} {\bibfnamefont {A.}~\bibnamefont
  {Mukhopadhyay}}, \bibinfo {author} {\bibfnamefont {X.-W.}\ \bibnamefont
  {Luo}}, \bibinfo {author} {\bibfnamefont {C.}~\bibnamefont {Schimelfenig}},
  \bibinfo {author} {\bibfnamefont {M.~K.~H.}\ \bibnamefont {Ome}}, \bibinfo
  {author} {\bibfnamefont {S.}~\bibnamefont {Mossman}}, \bibinfo {author}
  {\bibfnamefont {C.}~\bibnamefont {Zhang}},\ and\ \bibinfo {author}
  {\bibfnamefont {P.}~\bibnamefont {Engels}},\ }\bibfield  {title} {\bibinfo
  {title} {{Observation of Momentum Space Josephson Effects in Weakly Coupled
  Bose-Einstein Condensates}},\ }\href
  {https://doi.org/10.1103/PhysRevLett.132.233403} {\bibfield  {journal}
  {\bibinfo  {journal} {Phys. Rev. Lett.}\ }\textbf {\bibinfo {volume} {132}},\
  \bibinfo {pages} {233403} (\bibinfo {year} {2024})}\BibitemShut {NoStop}%
\bibitem [{\citenamefont {Jim\'enez-Garc\'{\i}a}\ \emph
  {et~al.}(2012)\citenamefont {Jim\'enez-Garc\'{\i}a}, \citenamefont {LeBlanc},
  \citenamefont {Williams}, \citenamefont {Beeler}, \citenamefont {Perry},\
  and\ \citenamefont {Spielman}}]{PhysRevLett.108.225303}%
  \BibitemOpen
  \bibfield  {author} {\bibinfo {author} {\bibfnamefont {K.}~\bibnamefont
  {Jim\'enez-Garc\'{\i}a}}, \bibinfo {author} {\bibfnamefont {L.~J.}\
  \bibnamefont {LeBlanc}}, \bibinfo {author} {\bibfnamefont {R.~A.}\
  \bibnamefont {Williams}}, \bibinfo {author} {\bibfnamefont {M.~C.}\
  \bibnamefont {Beeler}}, \bibinfo {author} {\bibfnamefont {A.~R.}\
  \bibnamefont {Perry}},\ and\ \bibinfo {author} {\bibfnamefont {I.~B.}\
  \bibnamefont {Spielman}},\ }\bibfield  {title} {\bibinfo {title} {{Peierls
  Substitution in an Engineered Lattice Potential}},\ }\href
  {https://doi.org/10.1103/PhysRevLett.108.225303} {\bibfield  {journal}
  {\bibinfo  {journal} {Phys. Rev. Lett.}\ }\textbf {\bibinfo {volume} {108}},\
  \bibinfo {pages} {225303} (\bibinfo {year} {2012})}\BibitemShut {NoStop}%
\bibitem [{\citenamefont {Cheuk}\ \emph {et~al.}(2012)\citenamefont {Cheuk},
  \citenamefont {Sommer}, \citenamefont {Hadzibabic}, \citenamefont {Yefsah},
  \citenamefont {Bakr},\ and\ \citenamefont
  {Zwierlein}}]{PhysRevLett.109.095302}%
  \BibitemOpen
  \bibfield  {author} {\bibinfo {author} {\bibfnamefont {L.~W.}\ \bibnamefont
  {Cheuk}}, \bibinfo {author} {\bibfnamefont {A.~T.}\ \bibnamefont {Sommer}},
  \bibinfo {author} {\bibfnamefont {Z.}~\bibnamefont {Hadzibabic}}, \bibinfo
  {author} {\bibfnamefont {T.}~\bibnamefont {Yefsah}}, \bibinfo {author}
  {\bibfnamefont {W.~S.}\ \bibnamefont {Bakr}},\ and\ \bibinfo {author}
  {\bibfnamefont {M.~W.}\ \bibnamefont {Zwierlein}},\ }\bibfield  {title}
  {\bibinfo {title} {{Spin-Injection Spectroscopy of a Spin-Orbit Coupled Fermi
  Gas}},\ }\href {https://doi.org/10.1103/PhysRevLett.109.095302} {\bibfield
  {journal} {\bibinfo  {journal} {Phys. Rev. Lett.}\ }\textbf {\bibinfo
  {volume} {109}},\ \bibinfo {pages} {095302} (\bibinfo {year}
  {2012})}\BibitemShut {NoStop}%
\bibitem [{\citenamefont {Lu}\ \emph {et~al.}(2016)\citenamefont {Lu},
  \citenamefont {Schemmer}, \citenamefont {Aycock}, \citenamefont {Genkina},
  \citenamefont {Sugawa},\ and\ \citenamefont
  {Spielman}}]{PhysRevLett.116.200402}%
  \BibitemOpen
  \bibfield  {author} {\bibinfo {author} {\bibfnamefont {H.-I.}\ \bibnamefont
  {Lu}}, \bibinfo {author} {\bibfnamefont {M.}~\bibnamefont {Schemmer}},
  \bibinfo {author} {\bibfnamefont {L.~M.}\ \bibnamefont {Aycock}}, \bibinfo
  {author} {\bibfnamefont {D.}~\bibnamefont {Genkina}}, \bibinfo {author}
  {\bibfnamefont {S.}~\bibnamefont {Sugawa}},\ and\ \bibinfo {author}
  {\bibfnamefont {I.~B.}\ \bibnamefont {Spielman}},\ }\bibfield  {title}
  {\bibinfo {title} {{Geometrical Pumping with a Bose-Einstein Condensate}},\
  }\href {https://doi.org/10.1103/PhysRevLett.116.200402} {\bibfield  {journal}
  {\bibinfo  {journal} {Phys. Rev. Lett.}\ }\textbf {\bibinfo {volume} {116}},\
  \bibinfo {pages} {200402} (\bibinfo {year} {2016})}\BibitemShut {NoStop}%
\bibitem [{\citenamefont {Lu}\ \emph {et~al.}(2022)\citenamefont {Lu},
  \citenamefont {Reid}, \citenamefont {Fritsch}, \citenamefont {Pi\~neiro},\
  and\ \citenamefont {Spielman}}]{PhysRevLett.129.040402}%
  \BibitemOpen
  \bibfield  {author} {\bibinfo {author} {\bibfnamefont {M.}~\bibnamefont
  {Lu}}, \bibinfo {author} {\bibfnamefont {G.~H.}\ \bibnamefont {Reid}},
  \bibinfo {author} {\bibfnamefont {A.~R.}\ \bibnamefont {Fritsch}}, \bibinfo
  {author} {\bibfnamefont {A.~M.}\ \bibnamefont {Pi\~neiro}},\ and\ \bibinfo
  {author} {\bibfnamefont {I.~B.}\ \bibnamefont {Spielman}},\ }\bibfield
  {title} {\bibinfo {title} {{Floquet Engineering Topological Dirac Bands}},\
  }\href {https://doi.org/10.1103/PhysRevLett.129.040402} {\bibfield  {journal}
  {\bibinfo  {journal} {Phys. Rev. Lett.}\ }\textbf {\bibinfo {volume} {129}},\
  \bibinfo {pages} {040402} (\bibinfo {year} {2022})}\BibitemShut {NoStop}%
\bibitem [{\citenamefont {Ome}\ \emph {et~al.}(2024)\citenamefont {Ome},
  \citenamefont {He}, \citenamefont {Mukhopadhyay}, \citenamefont {Crowell},
  \citenamefont {Mossman}, \citenamefont {Bersano}, \citenamefont {Zhang},\
  and\ \citenamefont {Engels}}]{ome2024galilean}%
  \BibitemOpen
  \bibfield  {author} {\bibinfo {author} {\bibfnamefont {M.}~\bibnamefont
  {Ome}}, \bibinfo {author} {\bibfnamefont {H.}~\bibnamefont {He}}, \bibinfo
  {author} {\bibfnamefont {A.}~\bibnamefont {Mukhopadhyay}}, \bibinfo {author}
  {\bibfnamefont {E.}~\bibnamefont {Crowell}}, \bibinfo {author} {\bibfnamefont
  {S.}~\bibnamefont {Mossman}}, \bibinfo {author} {\bibfnamefont
  {T.}~\bibnamefont {Bersano}}, \bibinfo {author} {\bibfnamefont
  {Y.}~\bibnamefont {Zhang}},\ and\ \bibinfo {author} {\bibfnamefont
  {P.}~\bibnamefont {Engels}},\ }\bibfield  {title} {\bibinfo {title}
  {{Galilean invariant dynamics in an emergent spin-orbit coupled Zeeman
  lattice}},\ }\href {https://www.nature.com/articles/s42005-023-01506-4}
  {\bibfield  {journal} {\bibinfo  {journal} {Commun. Phys.}\ }\textbf
  {\bibinfo {volume} {7}},\ \bibinfo {pages} {9} (\bibinfo {year}
  {2024})}\BibitemShut {NoStop}%
\bibitem [{\citenamefont {Luo}\ \emph {et~al.}(2015)\citenamefont {Luo},
  \citenamefont {Wu}, \citenamefont {Chen}, \citenamefont {Lu}, \citenamefont
  {Wang},\ and\ \citenamefont {You}}]{LuoXY_2015}%
  \BibitemOpen
  \bibfield  {author} {\bibinfo {author} {\bibfnamefont {X.}~\bibnamefont
  {Luo}}, \bibinfo {author} {\bibfnamefont {L.}~\bibnamefont {Wu}}, \bibinfo
  {author} {\bibfnamefont {J.}~\bibnamefont {Chen}}, \bibinfo {author}
  {\bibfnamefont {R.}~\bibnamefont {Lu}}, \bibinfo {author} {\bibfnamefont
  {R.}~\bibnamefont {Wang}},\ and\ \bibinfo {author} {\bibfnamefont
  {L.}~\bibnamefont {You}},\ }\bibfield  {title} {\bibinfo {title} {Generating
  an effective magnetic lattice for ultracold atoms},\ }\href
  {https://doi.org/10.1088/1367-2630/17/8/083048} {\bibfield  {journal}
  {\bibinfo  {journal} {New Journal of Physics}\ }\textbf {\bibinfo {volume}
  {17}},\ \bibinfo {pages} {083048} (\bibinfo {year} {2015})}\BibitemShut
  {NoStop}%
\bibitem [{\citenamefont {Su}\ \emph {et~al.}(2015)\citenamefont {Su},
  \citenamefont {Gou}, \citenamefont {Liu}, \citenamefont {Spielman},
  \citenamefont {Santos}, \citenamefont {Acus}, \citenamefont {Mekys},
  \citenamefont {Ruseckas},\ and\ \citenamefont {Juzeliūnas}}]{Suref_2015}%
  \BibitemOpen
  \bibfield  {author} {\bibinfo {author} {\bibfnamefont {S.-W.}\ \bibnamefont
  {Su}}, \bibinfo {author} {\bibfnamefont {S.-C.}\ \bibnamefont {Gou}},
  \bibinfo {author} {\bibfnamefont {I.-K.}\ \bibnamefont {Liu}}, \bibinfo
  {author} {\bibfnamefont {I.~B.}\ \bibnamefont {Spielman}}, \bibinfo {author}
  {\bibfnamefont {L.}~\bibnamefont {Santos}}, \bibinfo {author} {\bibfnamefont
  {A.}~\bibnamefont {Acus}}, \bibinfo {author} {\bibfnamefont {A.}~\bibnamefont
  {Mekys}}, \bibinfo {author} {\bibfnamefont {J.}~\bibnamefont {Ruseckas}},\
  and\ \bibinfo {author} {\bibfnamefont {G.}~\bibnamefont {Juzeliūnas}},\
  }\bibfield  {title} {\bibinfo {title} {Position-dependent spin–orbit
  coupling for ultracold atoms},\ }\href
  {https://doi.org/10.1088/1367-2630/17/3/033045} {\bibfield  {journal}
  {\bibinfo  {journal} {New Journal of Physics}\ }\textbf {\bibinfo {volume}
  {17}},\ \bibinfo {pages} {033045} (\bibinfo {year} {2015})}\BibitemShut
  {NoStop}%
\bibitem [{\citenamefont {Yu}\ \emph {et~al.}(2017)\citenamefont {Yu},
  \citenamefont {Xu},\ and\ \citenamefont {You}}]{PhysRevA.95.013409}%
  \BibitemOpen
  \bibfield  {author} {\bibinfo {author} {\bibfnamefont {J.}~\bibnamefont
  {Yu}}, \bibinfo {author} {\bibfnamefont {Z.-F.}\ \bibnamefont {Xu}},\ and\
  \bibinfo {author} {\bibfnamefont {L.}~\bibnamefont {You}},\ }\bibfield
  {title} {\bibinfo {title} {{Generating topological optical flux lattices for
  ultracold atoms by modulated Raman and radio-frequency couplings}},\ }\href
  {https://doi.org/10.1103/PhysRevA.95.013409} {\bibfield  {journal} {\bibinfo
  {journal} {Phys. Rev. A}\ }\textbf {\bibinfo {volume} {95}},\ \bibinfo
  {pages} {013409} (\bibinfo {year} {2017})}\BibitemShut {NoStop}%
\bibitem [{\citenamefont {Huang}\ \emph {et~al.}(2019)\citenamefont {Huang},
  \citenamefont {Zhuang}, \citenamefont {Zhou}, \citenamefont {Pu},
  \citenamefont {Guo},\ and\ \citenamefont {Gong}}]{PhysRevA.100.053606}%
  \BibitemOpen
  \bibfield  {author} {\bibinfo {author} {\bibfnamefont {Y.-X.}\ \bibnamefont
  {Huang}}, \bibinfo {author} {\bibfnamefont {W.~F.}\ \bibnamefont {Zhuang}},
  \bibinfo {author} {\bibfnamefont {X.-F.}\ \bibnamefont {Zhou}}, \bibinfo
  {author} {\bibfnamefont {H.}~\bibnamefont {Pu}}, \bibinfo {author}
  {\bibfnamefont {G.-C.}\ \bibnamefont {Guo}},\ and\ \bibinfo {author}
  {\bibfnamefont {M.}~\bibnamefont {Gong}},\ }\bibfield  {title} {\bibinfo
  {title} {{Bose-Einstein condensate in Bloch bands with an off-diagonal
  periodic potential}},\ }\href {https://doi.org/10.1103/PhysRevA.100.053606}
  {\bibfield  {journal} {\bibinfo  {journal} {Phys. Rev. A}\ }\textbf {\bibinfo
  {volume} {100}},\ \bibinfo {pages} {053606} (\bibinfo {year}
  {2019})}\BibitemShut {NoStop}%
\bibitem [{\citenamefont {Li}\ \emph {et~al.}(2018)\citenamefont {Li},
  \citenamefont {Ye}, \citenamefont {Chen}, \citenamefont {Kartashov},
  \citenamefont {Torner},\ and\ \citenamefont {Konotop}}]{PhysRevA.98.061601}%
  \BibitemOpen
  \bibfield  {author} {\bibinfo {author} {\bibfnamefont {C.}~\bibnamefont
  {Li}}, \bibinfo {author} {\bibfnamefont {F.}~\bibnamefont {Ye}}, \bibinfo
  {author} {\bibfnamefont {X.}~\bibnamefont {Chen}}, \bibinfo {author}
  {\bibfnamefont {Y.~V.}\ \bibnamefont {Kartashov}}, \bibinfo {author}
  {\bibfnamefont {L.}~\bibnamefont {Torner}},\ and\ \bibinfo {author}
  {\bibfnamefont {V.~V.}\ \bibnamefont {Konotop}},\ }\bibfield  {title}
  {\bibinfo {title} {Topological edge states in rashba-dresselhaus
  spin-orbit-coupled atoms in a zeeman lattice},\ }\href
  {https://doi.org/10.1103/PhysRevA.98.061601} {\bibfield  {journal} {\bibinfo
  {journal} {Phys. Rev. A}\ }\textbf {\bibinfo {volume} {98}},\ \bibinfo
  {pages} {061601} (\bibinfo {year} {2018})}\BibitemShut {NoStop}%
\bibitem [{\citenamefont {Zezyulin}\ and\ \citenamefont
  {Konotop}(2022)}]{PhysRevA.105.063323}%
  \BibitemOpen
  \bibfield  {author} {\bibinfo {author} {\bibfnamefont {D.~A.}\ \bibnamefont
  {Zezyulin}}\ and\ \bibinfo {author} {\bibfnamefont {V.~V.}\ \bibnamefont
  {Konotop}},\ }\bibfield  {title} {\bibinfo {title} {Localization of ultracold
  atoms in zeeman lattices with incommensurate spin-orbit coupling},\ }\href
  {https://doi.org/10.1103/PhysRevA.105.063323} {\bibfield  {journal} {\bibinfo
   {journal} {Phys. Rev. A}\ }\textbf {\bibinfo {volume} {105}},\ \bibinfo
  {pages} {063323} (\bibinfo {year} {2022})}\BibitemShut {NoStop}%
\bibitem [{\citenamefont {Fang}\ and\ \citenamefont
  {Lin}(2024)}]{PhysRevE.109.064219}%
  \BibitemOpen
  \bibfield  {author} {\bibinfo {author} {\bibfnamefont {P.}~\bibnamefont
  {Fang}}\ and\ \bibinfo {author} {\bibfnamefont {J.}~\bibnamefont {Lin}},\
  }\bibfield  {title} {\bibinfo {title} {Soliton in bose-einstein condensates
  with helicoidal spin-orbit coupling under a zeeman lattice},\ }\href
  {https://doi.org/10.1103/PhysRevE.109.064219} {\bibfield  {journal} {\bibinfo
   {journal} {Phys. Rev. E}\ }\textbf {\bibinfo {volume} {109}},\ \bibinfo
  {pages} {064219} (\bibinfo {year} {2024})}\BibitemShut {NoStop}%
\bibitem [{\citenamefont {Wu}\ \emph {et~al.}(2002)\citenamefont {Wu},
  \citenamefont {Diener},\ and\ \citenamefont {Niu}}]{PhysRevA.65.025601}%
  \BibitemOpen
  \bibfield  {author} {\bibinfo {author} {\bibfnamefont {B.}~\bibnamefont
  {Wu}}, \bibinfo {author} {\bibfnamefont {R.~B.}\ \bibnamefont {Diener}},\
  and\ \bibinfo {author} {\bibfnamefont {Q.}~\bibnamefont {Niu}},\ }\bibfield
  {title} {\bibinfo {title} {{Bloch waves and bloch bands of Bose-Einstein
  condensates in optical lattices}},\ }\href
  {https://doi.org/10.1103/PhysRevA.65.025601} {\bibfield  {journal} {\bibinfo
  {journal} {Phys. Rev. A}\ }\textbf {\bibinfo {volume} {65}},\ \bibinfo
  {pages} {025601} (\bibinfo {year} {2002})}\BibitemShut {NoStop}%
\bibitem [{\citenamefont {Hou}\ \emph {et~al.}(2018)\citenamefont {Hou},
  \citenamefont {Luo}, \citenamefont {Sun}, \citenamefont {Bersano},
  \citenamefont {Gokhroo}, \citenamefont {Mossman}, \citenamefont {Engels},\
  and\ \citenamefont {Zhang}}]{PhysRevLett.120.120401}%
  \BibitemOpen
  \bibfield  {author} {\bibinfo {author} {\bibfnamefont {J.}~\bibnamefont
  {Hou}}, \bibinfo {author} {\bibfnamefont {X.-W.}\ \bibnamefont {Luo}},
  \bibinfo {author} {\bibfnamefont {K.}~\bibnamefont {Sun}}, \bibinfo {author}
  {\bibfnamefont {T.}~\bibnamefont {Bersano}}, \bibinfo {author} {\bibfnamefont
  {V.}~\bibnamefont {Gokhroo}}, \bibinfo {author} {\bibfnamefont
  {S.}~\bibnamefont {Mossman}}, \bibinfo {author} {\bibfnamefont
  {P.}~\bibnamefont {Engels}},\ and\ \bibinfo {author} {\bibfnamefont
  {C.}~\bibnamefont {Zhang}},\ }\bibfield  {title} {\bibinfo {title}
  {{Momentum-Space Josephson Effects}},\ }\href
  {https://doi.org/10.1103/PhysRevLett.120.120401} {\bibfield  {journal}
  {\bibinfo  {journal} {Phys. Rev. Lett.}\ }\textbf {\bibinfo {volume} {120}},\
  \bibinfo {pages} {120401} (\bibinfo {year} {2018})}\BibitemShut {NoStop}%
\bibitem [{\citenamefont {Pitaevskii}\ and\ \citenamefont
  {Stringari}(2016)}]{pitaevskii2016bose}%
  \BibitemOpen
  \bibfield  {author} {\bibinfo {author} {\bibfnamefont {L.}~\bibnamefont
  {Pitaevskii}}\ and\ \bibinfo {author} {\bibfnamefont {S.}~\bibnamefont
  {Stringari}},\ }\href@noop {} {\emph {\bibinfo {title} {{Bose-Einstein
  condensation and superfluidity}}}},\ Vol.\ \bibinfo {volume} {164}\ (\bibinfo
   {publisher} {Oxford University Press},\ \bibinfo {year} {2016})\BibitemShut
  {NoStop}%
\bibitem [{\citenamefont {Chauveau}\ \emph {et~al.}(2023)\citenamefont
  {Chauveau}, \citenamefont {Maury}, \citenamefont {Rabec}, \citenamefont
  {Heintze}, \citenamefont {Brochier}, \citenamefont {Nascimbene},
  \citenamefont {Dalibard}, \citenamefont {Beugnon}, \citenamefont {Roccuzzo},\
  and\ \citenamefont {Stringari}}]{PhysRevLett.130.226003}%
  \BibitemOpen
  \bibfield  {author} {\bibinfo {author} {\bibfnamefont {G.}~\bibnamefont
  {Chauveau}}, \bibinfo {author} {\bibfnamefont {C.}~\bibnamefont {Maury}},
  \bibinfo {author} {\bibfnamefont {F.}~\bibnamefont {Rabec}}, \bibinfo
  {author} {\bibfnamefont {C.}~\bibnamefont {Heintze}}, \bibinfo {author}
  {\bibfnamefont {G.}~\bibnamefont {Brochier}}, \bibinfo {author}
  {\bibfnamefont {S.}~\bibnamefont {Nascimbene}}, \bibinfo {author}
  {\bibfnamefont {J.}~\bibnamefont {Dalibard}}, \bibinfo {author}
  {\bibfnamefont {J.}~\bibnamefont {Beugnon}}, \bibinfo {author} {\bibfnamefont
  {S.~M.}\ \bibnamefont {Roccuzzo}},\ and\ \bibinfo {author} {\bibfnamefont
  {S.}~\bibnamefont {Stringari}},\ }\bibfield  {title} {\bibinfo {title}
  {{Superfluid Fraction in an Interacting Spatially Modulated Bose-Einstein
  Condensate}},\ }\href {https://doi.org/10.1103/PhysRevLett.130.226003}
  {\bibfield  {journal} {\bibinfo  {journal} {Phys. Rev. Lett.}\ }\textbf
  {\bibinfo {volume} {130}},\ \bibinfo {pages} {226003} (\bibinfo {year}
  {2023})}\BibitemShut {NoStop}%
\bibitem [{\citenamefont {Tao}\ \emph {et~al.}(2023)\citenamefont {Tao},
  \citenamefont {Zhao},\ and\ \citenamefont
  {Spielman}}]{PhysRevLett.131.163401}%
  \BibitemOpen
  \bibfield  {author} {\bibinfo {author} {\bibfnamefont {J.}~\bibnamefont
  {Tao}}, \bibinfo {author} {\bibfnamefont {M.}~\bibnamefont {Zhao}},\ and\
  \bibinfo {author} {\bibfnamefont {I.~B.}\ \bibnamefont {Spielman}},\
  }\bibfield  {title} {\bibinfo {title} {{Observation of Anisotropic Superfluid
  Density in an Artificial Crystal}},\ }\href
  {https://doi.org/10.1103/PhysRevLett.131.163401} {\bibfield  {journal}
  {\bibinfo  {journal} {Phys. Rev. Lett.}\ }\textbf {\bibinfo {volume} {131}},\
  \bibinfo {pages} {163401} (\bibinfo {year} {2023})}\BibitemShut {NoStop}%
\bibitem [{\citenamefont {Leggett}(2006)}]{leggett2006quantum}%
  \BibitemOpen
  \bibfield  {author} {\bibinfo {author} {\bibfnamefont {A.~J.}\ \bibnamefont
  {Leggett}},\ }\href@noop {} {\emph {\bibinfo {title} {{Quantum liquids: Bose
  condensation and Cooper pairing in condensed-matter systems}}}}\ (\bibinfo
  {publisher} {Oxford university press},\ \bibinfo {year} {2006})\BibitemShut
  {NoStop}%
\bibitem [{\citenamefont {Biagioni}\ \emph {et~al.}(2024)\citenamefont
  {Biagioni}, \citenamefont {Antolini}, \citenamefont {Donelli}, \citenamefont
  {Pezz{\`e}}, \citenamefont {Smerzi}, \citenamefont {Fattori}, \citenamefont
  {Fioretti}, \citenamefont {Gabbanini}, \citenamefont {Inguscio},
  \citenamefont {Tanzi} \emph {et~al.}}]{biagioni2024measurement}%
  \BibitemOpen
  \bibfield  {author} {\bibinfo {author} {\bibfnamefont {G.}~\bibnamefont
  {Biagioni}}, \bibinfo {author} {\bibfnamefont {N.}~\bibnamefont {Antolini}},
  \bibinfo {author} {\bibfnamefont {B.}~\bibnamefont {Donelli}}, \bibinfo
  {author} {\bibfnamefont {L.}~\bibnamefont {Pezz{\`e}}}, \bibinfo {author}
  {\bibfnamefont {A.}~\bibnamefont {Smerzi}}, \bibinfo {author} {\bibfnamefont
  {M.}~\bibnamefont {Fattori}}, \bibinfo {author} {\bibfnamefont
  {A.}~\bibnamefont {Fioretti}}, \bibinfo {author} {\bibfnamefont
  {C.}~\bibnamefont {Gabbanini}}, \bibinfo {author} {\bibfnamefont
  {M.}~\bibnamefont {Inguscio}}, \bibinfo {author} {\bibfnamefont
  {L.}~\bibnamefont {Tanzi}}, \emph {et~al.},\ }\bibfield  {title} {\bibinfo
  {title} {{Measurement of the superfluid fraction of a supersolid by Josephson
  effect}},\ }\href {https://www.nature.com/articles/s41586-024-07361-9}
  {\bibfield  {journal} {\bibinfo  {journal} {Nature}\ }\textbf {\bibinfo
  {volume} {629}},\ \bibinfo {pages} {773} (\bibinfo {year}
  {2024})}\BibitemShut {NoStop}%
\bibitem [{\citenamefont {Wu}\ and\ \citenamefont
  {Niu}(2001)}]{PhysRevA.64.061603}%
  \BibitemOpen
  \bibfield  {author} {\bibinfo {author} {\bibfnamefont {B.}~\bibnamefont
  {Wu}}\ and\ \bibinfo {author} {\bibfnamefont {Q.}~\bibnamefont {Niu}},\
  }\bibfield  {title} {\bibinfo {title} {Landau and dynamical instabilities of
  the superflow of bose-einstein condensates in optical lattices},\ }\href
  {https://doi.org/10.1103/PhysRevA.64.061603} {\bibfield  {journal} {\bibinfo
  {journal} {Phys. Rev. A}\ }\textbf {\bibinfo {volume} {64}},\ \bibinfo
  {pages} {061603} (\bibinfo {year} {2001})}\BibitemShut {NoStop}%
\end{thebibliography}%

\end{document}